\documentclass{iopart}
\input epsf
\begin{document}
\title{Composite Fermions in Fractional Quantum Hall Systems}
\author{
   John J. Quinn}
\address{
   University of Tennessee, Knoxville, Tennessee 37996, USA}
\author{
   Arkadiusz W\'ojs}
\address{
   University of Tennessee, Knoxville, Tennessee 37996, USA\\
   and Wroclaw University of Technology, 50-370 Wroclaw, Poland}
\begin{abstract}
The composite Fermion (CF) picture offers a simple intuitive way 
of understanding many of the surprising properties of a strongly
interacting two-dimensional electron fluid in a large magnetic 
field.
The simple way in which the mean field CF picture describes the 
low lying bands of angular momentum multiplets for any value of 
the applied magnetic field is illustrated and compared with the 
results of exact numerical diagonalization of small systems.
The justification of the success of the CF approach is discussed
in some detail, and a CF hierarchy picture of the incompressible 
quantum fluid states is introduced.
The CF picture is used to understand the energy spectrum and
photoluminescence  of systems containing both electrons and valence
band holes.
\end{abstract}

\section{
   Introduction}
\label{sec1}
The study of the electronic properties of quasi-two-dimensional
(2D) systems has resulted in a number of remarkable discoveries 
in the past two decades \cite{ep2ds}.
Among the most interesting of these are the integral \cite{klitzing} 
and fractional \cite{tsui} quantum Hall effects.
In both of these effects, incompressible states of a 2D electron
liquid are found at particular values of the electron density for 
a given value of the magnetic field applied normal to the 2D layer.

The integral quantum Hall effect (IQHE) is rather simple to 
understand.
The incompressibility results from a cyclotron energy gap, 
$\hbar\omega_c$, in the single particle spectrum.
When all states below the gap are filled and all states above it
are empty, it takes a finite energy $\hbar\omega_c$ to produce
an infinitesimal compression.
Excited states consist of electron--hole pair excitations and 
require a finite excitation energy.
Both localized \cite{anderson} and extended single particle states 
are necessary to understand the experimentally observed behavior of 
the magneto-conductivity \cite{laughlin1}.

The fractional quantum Hall effect (FQHE) is more difficult to 
understand and more interesting in terms of new basic physics.
The energy gap that gives rise to the Laughlin \cite{laughlin2} 
incompressible fluid state is completely the result of the 
interaction between the electrons.
The elementary excitations are fractionally charged Laughlin 
quasiparticles, which satisfy fractional statistics \cite{halperin1}.
The standard techniques of many body perturbation theory are 
incapable of treating FQH systems because of the complete degeneracy 
of the single particle levels in the absence of the interactions.
Laughlin \cite{laughlin2} was able to determine the form of the ground 
state wave function and of the elementary excitations on the basis of
physical insight into the nature of the many body correlations.
Striking confirmation of Laughlin's picture was obtained by exact
diagonalization of the interaction Hamiltonian within the subspace
of the lowest Landau level of small systems \cite{haldane1}.
Jain \cite{jain1}, Lopez and Fradkin \cite{lopez}, and Halperin et al. 
 \cite{halperin2} have extended Laughlin's approach and developed 
a composite Fermion (CF) description of the 2D electron gas in 
a strong magnetic field.
This CF description has offered a simple picture for the 
interpretation of many experimental results.
However, the underlying reason for the validity of many of the 
approximations used with the CF approach is not completely 
understood \cite{wojs1}.

The object of this review is to present a simple and understandable
summary of the CF picture as applied to FQH systems.
Exact numerical calculations for up to eleven electrons on a spherical
surface will be compared with the predictions of the mean field CF
picture.
The CF hierarchy \cite{sitko1} will be introduced, and its predictions 
compared with numerical results.
It will be shown that sometimes the mean field CF hierarchy 
correctly predicts Laughlin-like incompressible ground states, and 
that sometimes it fails.

The CF hierarchy depends on the validity of the mean field 
approximation.
This seems to work well in predicting not only the Laughlin--Jain
families of incompressible ground states at particular values of
the applied magnetic field, but also in predicting the lowest lying
band of states at any value of the magnetic field.
The question of when the mean field CF picture works and why 
\cite{wojs1} will be discussed in some detail.
As first suggested by Haldane \cite{haldane1}, the behavior of the 
pseudopotential $V(L)$ describing the energy of interaction of 
a pair of electrons as a function of their total angular momentum 
$L$ is of critical importance.
Some examples of other strongly interacting 2D Fermion systems will
be presented, and some problems not yet completely understood will
be discussed.

The plan of the paper is as follows.
In section~\ref{sec2} the single particle states for electrons confined 
to a plane in the presence of an applied magnetic field \cite{gasiorowicz} 
are introduced.
The integral and fractional quantum Hall effects are discussed briefly.
Haldane's idea \cite{haldane1} that the condensation of Laughlin 
quasiparticles leads to a hierarchy containing all odd denominator 
fractions is discussed.
In section~\ref{sec3} the numerical calculations for a finite number of 
electrons confined to a spherical surface in the presence of a radial 
magnetic field are discussed.
Results for a ten electron system at different values of the magnetic 
field are presented.
In section~\ref{sec4} the ideas of fractional statistics and the 
Chern--Simons transformation are introduced.
In section~\ref{sec5} Jain's CF approach \cite{jain1} is outlined.
The sequence of Jain condensed states (given by filling factor
$\nu=n(1+2pn)^{-1}$, where $n$ is any integer and $p$ is a positive
integer) is shown to result from the mean field approximation.
The application of the CF picture to electrons on a spherical surface
is shown to predict the lowest band of angular momentum multiplets
in a very simple way that involves only the elementary problem of 
addition of angular momenta \cite{chen1}.
In section~\ref{sec6} the two energy scales, the Landau level separation 
$\hbar\omega_c$ and the Coulomb energy $e^2/\lambda$ (where 
$\lambda$ is the magnetic length), are discussed.
It is emphasized that the Coulomb interactions and Chern--Simons gauge
interactions between fluctuations (beyond the mean field) cannot 
possibly cancel for arbitrary values the applied magnetic field.
The reason for the success of the CF picture is discussed in terms 
of the behavior of the pseudopotential $V(L)$ and a kind of ``Hund's
rule'' for monopole harmonics \cite{wojs1}.
In section~\ref{sec7}, a phenomenological Fermi liquid picture is introduced
to describe low lying excited states containing three or more Laughlin
quasiparticles \cite{sitko2}.
In section~\ref{sec8} the CF hierarchy picture \cite{sitko1} is introduced.
Comparison with exact numerical results indicates that the behavior
of the quasiparticle pseudopotential is of critical importance in
determining the validity of this picture at a particular level of
the hierarchy.
In section~\ref{sec9} systems containing electrons and valence band 
holes are investigated \cite{wojs2}.
The photoluminescence and the role of excitons and negatively charged 
exciton complexes is discussed.
The final section is a summary.

\section{
   Integral and Fractional Quantum Hall Effects}
\label{sec2}
The Hamiltonian for an electron confined to the $x$--$y$ plane in the 
presence of a perpendicular magnetic field $\bi{B}$ is
\begin{equation}
   H_0={1\over2\mu}\left(\bi{p}+{e\over c}\bi{A}\right)^2.
\end{equation}
Here $\mu$ is the effective mass, $\bi{p}=(p_x,p_y,0)$ is the momentum 
operator and $\bi{A}(x,y)$ is the vector potential (whose curl gives 
$\bi{B}$).
For the ``symmetric gauge,'' $\bi{A}={1\over2}B(y,-x,0)$, the single
particle eigenfunctions \cite{gasiorowicz} are of the form $\psi_{nm}
(r,\theta)=e^{-im\theta}u_{nm}(r)$.
The angular momentum of the state $\psi_{nm}$ is $-m$ and its eigenenergy 
is given by
\begin{equation}
   E_{nm}={1\over2}\hbar\omega_c(2n+1+|m|-m).
\end{equation}
In these equations, $\omega_c=eB/\mu c$ is the cyclotron frequency,
$n=0$, 1, 2, \dots, and $m=0$, $\pm1$,  $\pm2$, \dots.
The lowest energy states (lowest Landau level) have $n=0$ and $m=0$, 
1, 2, \dots\ and energy $E_{0m}={1\over2}\hbar\omega_c$.
It is convenient to introduce a complex coordinate $z=re^{-i\theta}
=x-iy$, and to write the lowest Landau level wavefunctions as 
\begin{equation}
\label{eq3}
   \psi_{0m}(z)=N_mz^me^{-|z|^2/4}, 
\end{equation}
where $N_m$ is a normalization constant.
In this expression we have used the magnetic length $\lambda=
\sqrt{\hbar c/eB}$ as the unit of length.
The function $|\psi_{0m}|^2$ has its maximum value at a radius 
$r_m$ which is proportional to $\sqrt{m}$.
All single particle states belonging to a given Landau level are 
degenerate, and separated in energy from neighboring levels by
$\hbar\omega_c$.

If the system has a ``finite radial range,'' then the $m$ values 
are restricted to being less than some maximum value ($m=0$, 1, 
2, \dots, $N_\phi-1$).
The value of $N_\phi$ (the Landau level degeneracy) is equal to the 
total flux through the sample, $BC$ (where $C$ is the area), 
divided by the quantum of flux $\phi_0=hc/e$.
The filling factor $\nu$ is defined as the ratio of the number of 
electrons, $N$, to $N_\phi$.
When $\nu$ has an integral value, an infinitesimal decrease in the 
area $C$ requires promotion of an electron across the cyclotron gap 
$\hbar\omega_c$ to the first unoccupied Landau level, making the 
system incompressible.
This incompressibility together with the existence of both localized
and extended states in the system is responsible for the observed 
behavior of the magneto-conductivity of quantum Hall systems at 
integral filling factors \cite{laughlin1}.

In order to construct a many electron wavefunction $\Psi(z_1,z_2,\dots,
z_N)$ corresponding to a completely filled lowest Landau level, the 
product function which places one electron in each of the $N_\phi=N$ 
orbitals $\psi_{0m}$ ($m=0$, 1, \dots, $N_\phi-1$) must be antisymmetrized.
This can be done with the aid of a Slater determinant
\begin{equation}
\label{eq4}
   \Psi\propto\left| 
   \matrix{ 1 & 1 & \dots & 1 \cr
            z_1 & z_2 & \dots & z_N \cr
            z_1^2 & z_2^2 & \dots & z_N^2 \cr
            \vdots & \vdots & & \vdots \cr
            z_1^{N-1} & z_2^{N-1} & \dots & z_N^{N-1} }
   \right| \exp\left(-{1\over4}\sum_k|z_k|^2\right).
\end{equation}
The determinant in equation~(\ref{eq4}) is the well-known Vandemonde 
determinant.
It is not difficult to show that it is equal to $\prod_{i<j}(z_i-z_j)$.
Of course, $N_\phi$ is equal to $N$ (since each of the $N_\phi$ orbitals 
is occupied by one electron) and the filling factor $\nu=1$.

Laughlin noticed that if the factor $(z_i-z_j)$ arising from the
Vandemonde determinant was replaced by $(z_i-z_j)^{2p+1}$, where $p$ 
was an integer, the wavefunction 
\begin{equation}
\label{eq5}
   \Psi_{2p+1}
   \propto\prod_{i<j}(z_i-z_j)^{2p+1}
   \exp\left(-{1\over4}\sum_i|z_i|^2\right)
\end{equation}
would be antisymmetric, keep the electrons further apart (and therefore
reduce the Coulomb repulsion), and correspond to a filling factor 
$\nu=(2p+1)^{-1}$.
This results because the highest power of $z_i$ in the polynomial 
factor in $\Psi_{2p+1}$ is $(2p+1)(N-1)$ and it must be equal to the 
highest orbital index ($m=N_\phi-1$), giving $N_\phi-1=(2p+1)(N-1)$ and 
$\nu=N/N_\phi$ equal to $(2p+1)^{-1}$ in the limit of large systems.
The additional factor $\prod_{i<j}(z_i-z_j)^{2p}$ multiplying 
$\Psi_{m=1}$ is the Jastrow factor which accounts for correlations 
between electrons.

It is observed experimentally that states with filling factors $\nu=2/5$,
3/5, 3/7, etc.\ exhibit FQH behavior in addition to the Laughlin 
$\nu=(2p+1)^{-1}$ states.
Haldane \cite{haldane1} suggested that a hierarchy of condensed states 
arose from the condensation of quasiparticles (QP's) of ``parent'' 
FQH states.
In his picture, Laughlin condensed states of the electron system occurred
when $N_\phi=(2p+1)N_e$, where the exponent $2p+1$ in equation~(\ref{eq5}) 
was an odd integer and the symbol $N_e$ denoted the number of electrons.
Condensed QP states occurred when $N_e=2qN_{\rm QP}$, because the number 
of places available for inserting a QP in a Laughlin state was $N_e$.
Haldane required the exponent $2q$ to be even ``because the QP's are 
bosons.''
This scheme gives rise to a hierarchy of condensed states which contains 
all odd denominator fractions.
Haldane cautioned that the validity of the hierarchy scheme at a particular
level depended upon the QP interactions which were totally unknown.

\section{
   Numerical Study of Small Systems}
\label{sec3}
Haldane \cite{haldane1} introduced the idea of putting a small number of 
electrons on a spherical surface of radius $R$ at the center of which is 
a magnetic monopole of strength $2S\phi_0$.
The single particle Hamiltonian can be expressed as \cite{fano}
\begin{equation}
   H_0={\hbar^2\over2\mu R^2}(\bi{L}-S\hat{R})^2,
\end{equation}
where $\bi{L}$ is the angular momentum operator (in units of $\hbar$),
$\hat{R}$ is the unit vector in the radial direction, and $\mu$ is 
the mass.
The components of $\bi{L}$ satisfy the usual commutation rules 
$[L_\alpha,L_\beta]=i\epsilon_{\alpha\beta\gamma}L_\gamma$.
The eigenstates of $H_0$ can be denoted by $\left|l,m\right>$;
they are  eigenfunctions of $L^2$ and $L_z$ with eigenvalues $l(l+1)$ 
and $m$, respectively.
The lowest energy eigenvalue (shell) occurs for $l=S$ and has energy 
${1\over2}\hbar\omega_c$.
The $n$th excited shell has $l=S+n$, and
\begin{equation}
   E_n={\hbar\omega_c\over2S}\left[l(l+1)-S^2\right] 
      =\hbar\omega_c\left[n+{1\over2}+{n(n+1)\over2S}\right],
\end{equation}
where the cyclotron energy is equal to $\hbar\omega_c=S\hbar^2/\mu 
R^2$ and the magnetic length is $\lambda=R/\sqrt{S}$.
If we concentrate on a partially filled lowest Landau level we have 
only $N_\phi=2S+1$ degenerate single particle states (since the electron
angular momentum $l$ must be equal to $S$ and its $z$-component $m$
can take on values between $-l$ and $l$).
The Hilbert space ${\cal H}_{\rm MB}$ of $N$ electrons in these $N_\phi$ 
single particle states contains $N_{\rm MB}=N_\phi![N!(N_\phi-N)!]^{-1}$ 
antisymmetric many body states.
The single particle configurations $\left|m_1,m_2,\dots,m_N\right>=
c_{m_1}^\dagger c_{m_2}^\dagger\dots c_{m_N}^\dagger\left|{\rm vac}
\right>$ can be chosen as a basis of ${\cal H}_{\rm MB}$.
Here $c_m^\dagger$ creates an electron in the single particle state 
$\left|l=S,m\right>$, and $\left|{\rm vac}\right>$ is the vacuum state.
The space ${\cal H}_{\rm MB}$ can also be spanned by the angular 
momentum eigenfunctions, $\left|L,M,\alpha\right>$, where $L$ is the 
total angular momentum, $M$ its $z$-component, and $\alpha$ is a label 
which distinguishes different multiplets with the same $L$.
If $\hbar\omega_c\gg e^2/\lambda$, the diagonalization of the interaction 
Hamiltonian
\begin{equation}
   H_I=\sum_{i<j}{e^2\over r_{ij}}
\end{equation}
in the Hilbert space ${\cal H}_{\rm MB}$ of the lowest Landau level
gives an excellent approximation to exact eigenstates of an interacting
$N$ electron system.
The single particle configuration basis is particularly convenient 
since the many body interaction matrix elements in this basis, 
$\left<m_1,m_2,\dots,m_N|H_I|m'_1,m'_2,\dots,m'_N\right>$, are 
expressed through the two body ones, $\left<m_1,m_2|H_I|m'_1,m'_2
\right>$, in a very simple way.
On the other hand, using the angular momentum eigenstates $\left|L,M,
\alpha\right>$ allows the explicit decomposition of the total Hilbert 
space ${\cal H}_{\rm MB}$ into total angular momentum eigensubspaces.
Because the interaction Hamiltonian is a scalar, the Wigner--Eckart
theorem tells us that
\begin{equation}
   \left<L',M',\alpha'|H_I|L,M,\alpha\right>=
   \delta_{LL'}\delta_{MM'}V_{\alpha\alpha'}(L),
\end{equation}
where the reduced matrix element 
\begin{equation}
   V_{\alpha\alpha'}(L)=\left<L,\alpha'|H_I|L,\alpha\right>
\end{equation}
is independent of $M$.
The eigenfunctions of $L$ are simpler to find than those of $H_I$, 
because efficient numerical techniques exist for obtaining 
eigenfunctions of operators with known eigenvalues.
Finding the eigenfunctions of $L$ and then using the Wigner--Eckart 
theorem considerably reduces dimensions of the matrices that must be 
diagonalized to obtain eigenvalues of $H_I$.
Some matrix dimensions are listed in table~\ref{tab1}, where the 
degeneracy of the lowest Landau level and the dimensions of the total 
many body Hilbert space, $N_{\rm MB}$, and of the largest $M$ subspace, 
$N_{\rm MB}(M=0)$, are given for the Laughlin $\nu=1/3$ state of six 
to eleven electron systems (the $N$ electron Laughlin $\nu=(2p+1)^{-1}$ 
state occurs at $N_\phi=(2p+1)(N-1)$).
\begin{table}
\caption{
   The Landau level degeneracy $N_\phi=2S+1$ and the dimensions of 
   the total $N$ electron Hilbert space, $N_{\rm MB}$, and of the 
   largest $M$ subspace, $N_{\rm MB}(M=0)$, at the filling factor 
   $\nu=1/3$.}
\begin{indented}
\item[]\begin{tabular}{@{}rcrrr}
\br
  $N$ & 
  $N_\phi$ & 
  $N_{\rm MB}$ & 
  $N_{\rm MB}(M=0)$ \\
\mr
    6 &   16 &         8,008 &       338  \\ 
    7 &   19 &        50,388 &     1,656  \\ 
    8 &   22 &       319,770 &     8,512  \\ 
    9 &   25 &     2,042,975 &    45,207  \\ 
   10 &   28 &    13,123,110 &   246,448  \\
   11 &   31 &    84,672,315 & 1,371,535 \\
\br
\end{tabular}
\end{indented}
\label{tab1}
\end{table}
For example, in the eleven electron system at $\nu=1/3$, the $L=0$
block that must be diagonalized to obtain the Laughlin ground state 
is only 1160 by 1160, small compared to the total dimension of 
1,371,535 for the entire $M=0$ subspace.

\begin{figure}[t]
\rule{1in}{0in}
\epsfxsize=3.75in
\epsfbox{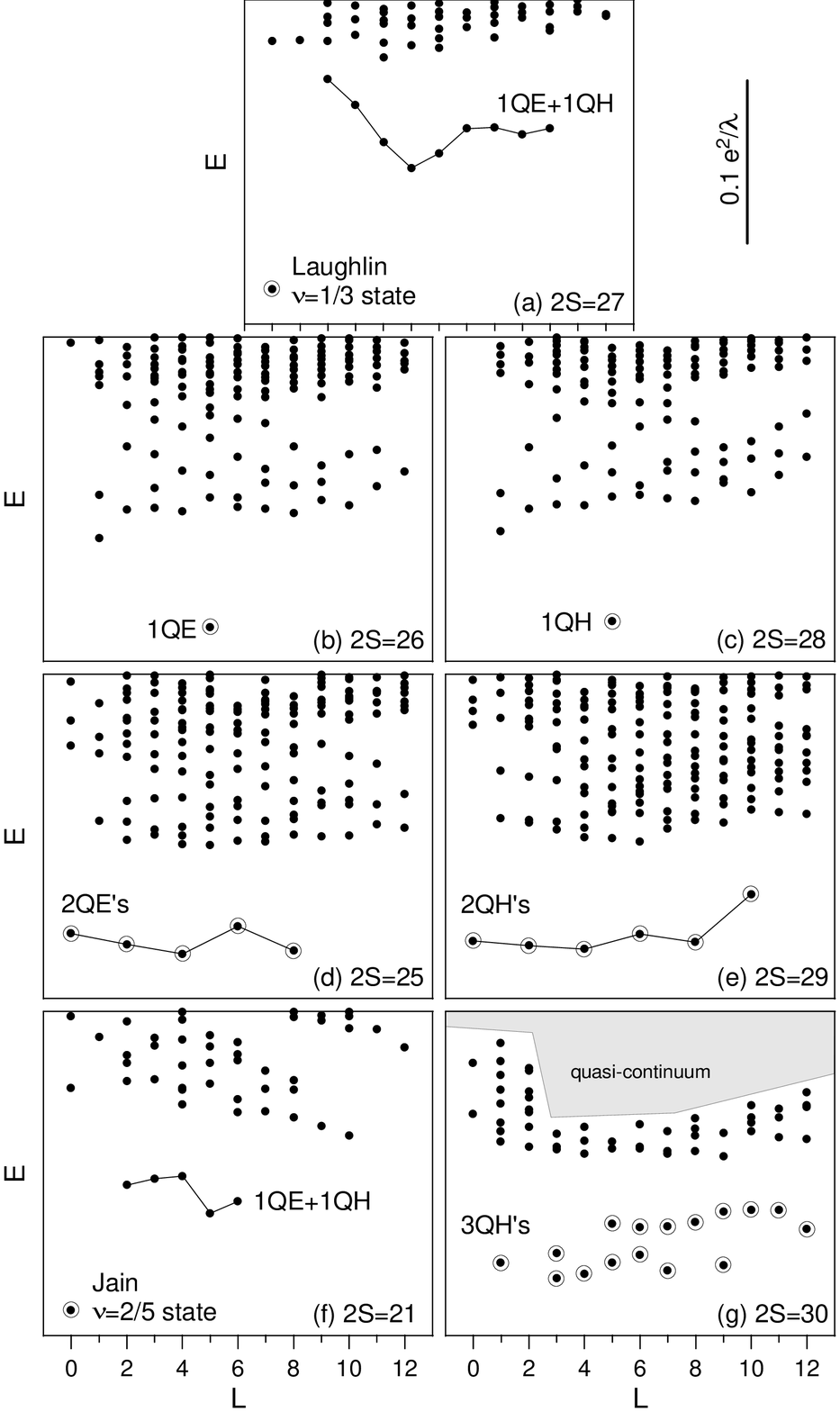}
\caption{
   The energy spectra of ten electrons in the lowest Landau level 
   calculated on a Haldane sphere with $2S$ between 21 and 30. 
   The open circles and solid lines mark lowest energy bands with 
   fewest composite Fermion quasiparticles.}
\label{fig1}
\end{figure}
Typical results for the energy spectrum are shown in figure~\ref{fig1}
for $N=10$ and a few different values of $2S$ between 21 and 30.
The low energy bands marked with open circles and solid lines will
be discussed in detail in the following sections.
Frames (a) and (f) show two $L=0$ incompressible ground states:
Laughlin state at $\nu=1/3$ and Jain state at $\nu=2/5$, respectively.
In other frames, a number of QP's form the lowest energy bands.

\section{
   Chern--Simons Transformation and Statistics in 2D Systems}
\label{sec4}
Before discussing the Chern--Simons gauge transformation and its 
relation to particle statistics, it is useful to look at a system 
of two particles each of charge $-e$ and mass $\mu$, confined to 
a plane, in the presence of a perpendicular magnetic field $\bi{B}
=(0,0,B)=\nabla\times\bi{A}(\bi{r})$.
Because $\bi{A}$ is linear in the coordinate $\bi{r}=(x,y)$
[e.g., in the symmetric gauge, $\bi{A}(\bi{r})={1\over2}B(y,-x)$],
the Hamiltonian separates into the center of mass (CM) and relative 
(REL) coordinate pieces, with  $\bi{R}={1\over2}(\bi{r}_1+\bi{r}_2)$ 
and $\bi{r}=\bi{r}_1-\bi{r}_2$ being the CM and REL coordinates, 
respectively.
The energy spectra of $H_{\rm CM}$ and $H_{\rm REL}$ are identical 
to that of a single particle of mass $\mu$ and charge $-e$.
We have already seen that for the lowest Landau level $\psi_{0m}=
N_mr^me^{-im\phi}e^{-r^2/4\lambda^2}$.
For the relative motion $\phi=\phi_1-\phi_2$, and an interchange 
of the pair, $P\psi(\bi{r}_1,\bi{r}_2)=\psi(\bi{r}_2,\bi{r}_1)$,
is accomplished by replacing $\phi$ by $\phi+\pi$.
In 3D systems, where two consecutive interchanges must result in the 
original wavefunction, this implies that $e^{im\pi}$ must be equal 
to either $+1$ ($m$ even; Bosons) or $-1$ ($m$ odd; Fermions).
It is well-known \cite{leinaas} that for 2D systems $m$ need not be 
an integer.
Interchange of a pair of identical particles can give $P\psi(\bi{r}_1,
\bi{r}_2)=e^{i\pi\theta}\psi(\bi{r}_1,\bi{r}_2)$, where the 
statistical parameter $\theta$ can assume non-integral values leading 
to anyon statistics.

A Chern--Simons (CS) transformation is a singular gauge transformation 
\cite{halperin2} in which an electron creation operator 
$\psi_e^\dagger(\bi{r})$ is replaced by a composite particle operator 
$\psi^\dagger(\bi{r})$ given by
\begin{equation}
   \psi^\dagger(\bi{r})=\psi_e^\dagger(\bi{r})\exp\left[ 
   i\alpha\int d^2\bi{r}' \arg(\bi{r}-\bi{r}')
   \psi^\dagger(\bi{r}')\psi(\bi{r}')\right].
\end{equation}
Here $\arg(\bi{r}-\bi{r}')$ is the angle the vector $\bi{r}-\bi{r}'$
makes with the $x$-axis and $\alpha$ is an arbitrary parameter. 
The kinetic energy operator can be written in terms of the transformed
operator as
\begin{equation}
   K={1\over2\mu}\int d^2\bi{r}\; \psi^\dagger(\bi{r})\left[
   -i\hbar\nabla
   +{e\over c}\bi{A}(\bi{r})
   +{e\over c}\bi{a}(\bi{r})
   \right]^2 \psi(\bi{r}).
\end{equation}
Here
\begin{equation}
\label{eq9a}
   \bi{a}_{\bi{r}'}(\bi{r})
   ={\alpha\phi_0\over2\pi}\cdot
   {\hat{z}\times(\bi{r}-\bi{r}')\over|\bi{r}-\bi{r}'|^2}
\end{equation}
and
\begin{equation}
\label{eq9}
   \bi{a}(\bi{r})=\alpha\phi_0\int d^2\bi{r}'\;
   \bi{a}_{\bi{r}'}(\bi{r})\;
   \psi^\dagger(\bi{r}')\psi(\bi{r}'),
\end{equation}
where $\hat{z}$ is a unit vector perpendicular to the 2D layer.
The CS transformation can be thought of as an attachment to each 
particle of flux tube carrying a fictitious flux $\alpha\phi_0$
(where $\phi_0=hc/e$ is the quantum of flux) and a fictitious charge
$-e$ which couples in the standard way to the vector potential 
caused by the flux tubes on every other particle.
The $\bi{a}_{\bi{r}'}(\bi{r})$ is interpreted as the vector 
potential at position $\bi{r}$ due to a magnetic flux of strength 
$\alpha\phi_0$ localized at $\bi{r}'$, and $\bi{a}(\bi{r})$ is 
the total vector potential at position $\bi{r}$ due to all CS 
fluxes.
The CS magnetic field associated with the particle at $\bi{r}'$ 
is $\bi{b}(\bi{r})=\nabla\times\bi{a}_{\bi{r}'}(\bi{r})=\alpha
\phi_0\delta(\bi{r}-\bi{r}')\hat{z}$.
Because two charged particles cannot occupy the same position, one 
particle never senses the magnetic field of other particles, but it 
does sense the vector potential resulting from their CS fluxes.
The classical equations of motion are unchanged by the presence of 
the CS flux, but the quantum statistics of the particles are changed
unless $\alpha$ is an even integer.

For the two particle system, the vector potential associated with 
the CS flux $\bi{a}_{\bi{r}_2}(\bi{r_1})$ depends only on the 
relative coordinate $\bi{r}=\bi{r}_1-\bi{r}_2$.
When $\bi{a}(\bi{r})$ is added to $\bi{A}(\bi{r})$, the vector 
potential of the applied magnetic field, the Schr\"odinger equation
has a solution 
\begin{equation}
   \tilde{\psi}_m=e^{-i\alpha\phi}\psi_m, 
\end{equation}
where $\psi_m$ is the solution with $\alpha=0$ (i.e. in the absence 
of CS flux).
If $\alpha$ is an odd integer, Boson and Fermion statistics are 
interchanged; if $\alpha$ is even, no change in statistics occurs and 
electrons are transformed into composite Fermions with an identical 
energy spectrum.

The Hamiltonian for the composite particle system (charged particles 
with attached flux tubes) is much more complicated than the original
system with $\alpha=0$.
What is gained by making the CS transformation?
The answer is that one can use the ``mean field'' approximation
in which $\bi{A}(\bi{r})+\bi{a}(\bi{r})$, the vector potential of
the external plus CS magnetic fields, is replaced by $\bi{A}(\bi{r})
+\left<\bi{a}(\bi{r})\right>$, where $\left<\bi{a}(\bi{r})\right>$
is the mean field value of $\bi{a}(\bi{r})$ obtained by simply
replacing $\varrho(\bi{r}')=\psi^\dagger(\bi{r}')\psi(\bi{r}')$ 
by its average value $\varrho_0$ in equation~(\ref{eq9}).
A mean field energy spectrum can be constructed in which the massive 
degeneracy of the original partially filled electron Landau level 
disappears.
One might then hope to treat both the Coulomb interaction and the CS 
gauge field interactions among the fluctuations (beyond the mean field)
by standard many body perturbation techniques (e.g. by the random phase
approximation, RPA).
Unfortunately, there is no small parameter for a many body perturbation
expansion unless $\alpha$, the number of CS flux quanta attached to each
particle, is small compared to unity.
However, a Landau--Silin \cite{silin} type Fermi liquid approach can 
take account of the short range correlations phenomenologically.
A number of excellent papers on anyon superconductivity \cite{laughlin3} 
treat CS gauge interactions by standard many body techniques.
Halperin and collaborators \cite{halperin2} have treated the half filled 
Landau level as a liquid of composite Fermions moving in zero effective 
magnetic field.
Their RPA--Fermi-liquid approach gives a surprisingly satisfactory
account of the properties of that state.

The vector potential associated with fluctuations beyond the mean field 
is given by $\delta\bi{a}(\bi{r})=\bi{a}(\bi{r})-\left<\bi{a}(\bi{r})
\right>$.
The perturbation to the mean field Hamiltonian contains both linear
and quadratic terms in $\delta\bi{a}(\bi{r})$, resulting in both
two body -- containing $\varrho(\bi{r}_1)\varrho(\bi{r}_2)$ --
and three body -- containing $\varrho(\bi{r}_1)\varrho(\bi{r}_2)
\varrho(\bi{r}_3)$ -- interaction terms.
The three body interaction terms are usually ignored, though for 
$\alpha$ of the order of unity this approximation is of questionable 
validity.

\section{
   Jain's Composite Fermion Picture}
\label{sec5}
Jain noted that in the mean field approximation, an effective filling 
factor $\nu^*$ of the composite Fermions was related to the electron 
filling factor $\nu$ by the relation
\begin{equation}
\label{eq10}
   (\nu^*)^{-1}=\nu^{-1}-2p.
\end{equation}
Remember that $\nu^{-1}$ is equal to the number of flux quanta of the 
applied magnetic field per electron, and $2p$ is the (even) number of 
CS flux quanta (oriented opposite to the applied magnetic field) 
attached to each electron in the CS transformation.
Equation~(\ref{eq10}) implies that when $\nu^*=\pm1$, $\pm2$, \dots\ 
(negative values correspond to the effective magnetic field $\bi{B}^*$ 
seen by the CF's oriented opposite to $\bi{B}$) and a non-degenerate 
mean field CF ground state occurs, then $\nu=\nu^*(1+2p\nu^*)^{-1}$.
This Jain sequence of condensed states ($\nu=1/3$, $2/5$, $3/7$, \dots\
and $\nu=2/3$, $3/5$, \dots\ for $p=1$) is the set of FQH states most 
prominent in experiment.
When $\nu^*$ is not an integer, QP's of the neighboring Jain state will 
occur.

It is quite remarkable that the mean field CF picture predicts not only 
the Jain sequence of incompressible ground states, but the correct band 
of low energy states for any value of the applied magnetic field.
This is very nicely illustrated for the case of $N$ electrons on a 
Haldane sphere.
When the monopole strength seen by an electron has the value $2S$,
the effective monopole strength seen by a CF is $2S^*=2S-2p(N-1)$.
This equation reflects the fact that a given CF senses the vector 
potential produced by the CS flux on all other particles, but not 
its own CS flux.
In table~\ref{tab2} the ten particle system is described for a number 
of values of $2S$ between 29 and 15.
\begin{table}
\caption{
   The effective CF monopole strength $2S^*$, the numbers of CF 
   quasiparticles (quasielectrons, $N_{\rm QE}$, and quasiholes, 
   $N_{\rm QH}$), the angular momentum of the lowest CF shell 
   $l^*$, the CF and electron filling factors $\nu^*$ and $\nu$, 
   and the angular momenta $L$ of the lowest lying band of 
   multiplets for a ten electron system at $2S$ between 29 and 15.}
\begin{indented}
\item[]\begin{tabular}{@{}llllllll}
\br
  $2S$         & 29 & 28 & 27 & 26 & 25 & 21 & 15 \\
\mr
  $2S^*$       & 11 & 10 &  9 &  8 &  7 &  3 & -3 \\
  $N_{\rm QH}$ &  2 &  1 &  0 &  0 &  0 &  0 &  0 \\
  $N_{\rm QE}$ &  0 &  0 &  0 &  1 &  2 &  6 &  6 \\
  $l_0^*$      &11/2&  5 & 9/2&  4 & 7/2& 3/2& 3/2\\
  $\nu^*$      &    &    &  1 &    &    &  2 & -2 \\
  $\nu$        &    &    & 1/3&    &    & 2/5& 2/3\\
  $L$          &0, 2, 4, 6, 8, 10& 5 & 0 & 5 & 0, 2, 4, 6, 8 & 0 & 0 \\
\br
\end{tabular}
\end{indented}
\label{tab2}
\end{table}
The Laughlin $\nu=1/3$ state occurs at $2S_3=3(N-1)=27$.
For values of $2S$ different from this value, $2S-2S_3=\pm N_{\rm QP}$ 
(``$+$'' corresponds to quasiholes, QH, and ``$-$'' to quasielectrons, 
QE).
Let us apply the CF description to the ten electron spectra in 
figure~\ref{fig1}.
At $2S=27$, we take $p=1$ and attach two CS flux quanta each electron.
This gives $2S^*=9$ so that the ten CF's completely fill the $2S^*+1$ 
states in the lowest angular momentum shell (lowest Landau level).
There is a gap $\hbar\omega_c^*=\hbar eB^*/\mu c$ to the next shell, 
which is responsible for the incompressibility of the Laughlin state.
Just as $|S|$ played the role of the angular momentum of the lowest 
shell of electrons, $l^*=|S^*|$ plays the role of the CF angular 
momentum and $2|S^*|+1$ is the degeneracy of the CF shell.
Thus, the states with $2S=26$ and 28 contain a single quasielectron
(QE) and quasihole (QH), respectively.
For the QE state, $2S^*=8$ and the lowest shell of angular momentum 
$l_0^*=4$ can accommodate only nine CF's.
The tenth is the QE in the $l_1^*=l_0^*+1=5$ shell, giving the total 
angular momentum $L=5$.
For the QH state, $2S^*=10$ and the lowest shell can accommodate 
eleven CF's each with angular momentum $l_0^*=5$.
The one empty state (QH) gives $L=l^*=5$.
For $2S=25$ we obtain $2S^*=7$, and there are two QE's each of 
angular momentum $l_1^*=9/2$ in the first excited CF shell.
Adding the angular momenta of the two QE's gives the band of 
multiplets $L=0$, 2, 4, 6, and 8.
Similarly, for $2S=29$ we obtain $2S^*=11$, and there are two QH's 
each with $l_0^*=11/2$, resulting in the allowed pair states at
$L=0$, 2, 4, 6, 8, and 10.
At $2S=21$, the lowest shell with $l_0^*=3/2$ can accommodate only 
four CF's, but the other six CF's exactly fill the excited $l_1^*=5/2$ 
shell.
The resulting incompressible ground state is the Jain $\nu=2/5$ state, 
since $\nu^*=2$ for the two filled shells.
A similar argument leads to $\nu^*=-2$ (minus sign means $\bi{B}^*$ 
oriented opposite to $\bi{B}$) and $\nu=2/3$ at $2S=15$.
At $2S=30$, the addition of three QH angular momenta of $l_0^*=6$ 
gives the following band of low lying multiplets $L=1$, $3^2$, 4, 
$5^2$, $6^2$, $7^2$, 8, $9^2$, 10, 11, 12, 13, and 15.
As demonstrated on an example in figure~\ref{fig1}, this simple mean 
field CF picture correctly predicts the band of low energy multiplets 
for any number of electrons $N$ and for any value of $2S$.

\section{
   Energy Scales and the Electron Pseudopotentials}
\label{sec6}
\begin{figure}[t]
\rule{1in}{0in}
\epsfxsize=3.75in
\epsffile{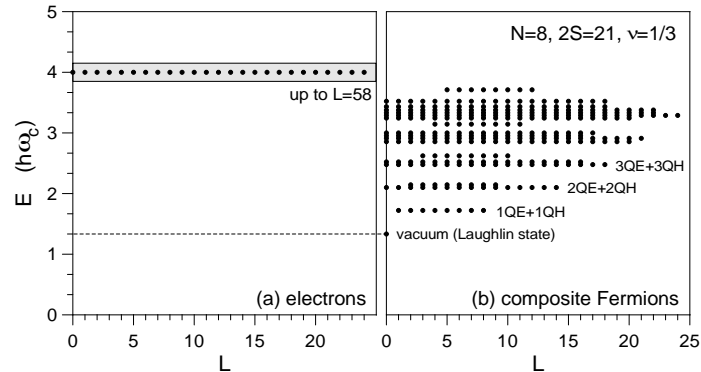}
\caption{
   The energy spectra of eight
   (a) non-interacting electrons and 
   (b) non-interacting composite Fermions. 
   The characteristic energy of the Coulomb interaction is marked
   in frame (a) with a shaded rectangle.}
\label{fig2}
\end{figure}

The mean field composite Fermion picture is remarkably successful in
predicting the low energy multiplets in the spectrum of $N$ electrons
on a Haldane sphere.
It was suggested originally that this success resulted from the 
cancellation of the Coulomb and Chern--Simons gauge interactions
among fluctuations beyond the mean field.
In figure~\ref{fig2}, we show the lowest bands of multiplets for eight
non-interacting electrons and for the same number of non-interacting
mean field CF's at $2S=21$.
The energy scale associated with the CS gauge interactions which 
convert the electron system in frame (a) to the CF system in frame 
(b) is $\hbar\omega_c^*\propto B$.
The energy scale associated with the electron-electron Coulomb 
interaction is $e^2/\lambda\propto\sqrt{B}$.
The Coulomb interaction lifts the degeneracy of the non-interacting 
electron bands in frame (a).
However, for very large value of $B$ the Coulomb energy can be made
arbitrarily small compared to the CS energy (as marked with a shaded 
rectangle in figure~\ref{fig2}), i.e. to the separation between the
CF Landau levels.
The energy separations in the mean field CF model are completely wrong, 
but the structure of the low lying states (i.e., which angular momentum 
multiplets form the low lying bands) is very similar to that of the
fully interacting electron system and completely different from that
of the non-interacting electron system.

\subsection{
   Two Fermion Problem}
An intuitive picture of why this occurs can be obtained by considering
the two Fermion problem.
The relative (REL) motion of a pair of electrons $(ij)$ is described 
by a coordinate $z_{ij}=z_i-z_j=r_{ij}e^{-i\phi_{ij}}$, and for the 
lowest Landau level its wavefunction contains a factor $z_{ij}^m$, 
where $m=1$, 3, 5, \dots.
If every pair of particles has identical behavior, the many particle
wavefunction must contain a similar factor for each pair giving 
a total factor $\prod_{i<j}z_{ij}^m$.
As we have seen, the highest power of $z_i$ in this product is 
$m(N-1)$.
If $m(N-1)$ is equal to $N_\phi-1=2S$, the maximum value of the 
$z$-component of the single particle angular momentum, the Laughlin
$\nu=m^{-1}$ wavefunction results.
For electrons, the $m$th cyclotron orbit, whose radius is $r_m$, 
encloses a flux $m\phi_0$ (i.e. $\pi r_m^2B=m\phi_0$).
For a Laughlin $\nu=m^{-1}$ state the pair function must have a 
radius $r_m=r_1\sqrt{m}$.
Let us describe the CF orbits by radius $\varrho_{\tilde{m}}$ and 
require that the $\tilde{m}$th orbit enclose $\tilde{m}$ flux quanta.
It is apparent that if a flux tube carrying two flux quanta (oriented
opposite to the applied magnetic field $B$) is attached to each electron 
in the CS transformation of the $\nu=1/3$ state, the smallest orbit of
radius $\varrho_{\tilde{m}=1}$ has exactly the same size as $r_{m=3}$.
Both orbits enclose three flux quanta of the applied field, but the CF 
orbit also encloses the two oppositely oriented CS flux quanta attached 
to the electrons to form the CF's.
In the absence of electron--electron interactions, the energies of these
orbits are unchanged, since they still belong to the degenerate single 
particle states of the lowest Landau level.

In  the mean field approximation the CS fluxes are replaced by a 
spatially uniform magnetic field, leading to an effective field $B^*=B/m$.
The orbits for the CF pair states in the mean field approximation are 
exactly the same as those of the exact CS Hamiltonian.
The smallest orbit has radius $\varrho_{\tilde{m}=1}$ equivalent to the 
electron orbit $r_{m=3}$.
However, in the mean field approximation, the energies are changed 
(because $\omega_c^*=eB^*/\mu c$ replaces $\omega_c$).
This energy change leads to completely incorrect mean field CF energies, 
but the mean field CF orbitals give the correct structure to the low 
lying set of multiplets.

In the presence of a repulsive interaction, the low lying energy 
states will have the largest possible value of $m$.
For a monopole strength $2S=m(N-1)$, where $m$ is an odd integer,
every pair can have radius $r_m$ and avoid the large repulsion
associated with $r_1$, $r_3$, \dots, $r_{m-2}$.
These ideas can be made somewhat more rigorous by using methods 
of atomic and nuclear physics for studying angular momentum shells
of interacting Fermions.

\subsection{
   Two Body Interaction Pseudopotential}
As first suggested by Haldane \cite{haldane1}, the behavior of the 
interacting many electron system depends entirely on the behavior 
of the two body interaction pseudopotential, which is defined as 
the interaction energy $V$ of a pair of electrons as a function of 
their pair angular momentum.
In the spherical geometry, in order to allow for meaningful comparison 
of the pseudopotentials obtained for different values of $2S$ (and thus 
different single electron angular momenta $l$), it is convenient to use 
the ``relative'' angular momentum ${\cal R}=2l-L_{12}$ rather than 
$L_{12}$ (the length of $\hat\bi{L}_{12}=\hat\bi{l}_1+\hat\bi{l}_2$).
The pair states with a given ${\cal R}=m$ (an odd integer) obtained 
on a sphere for different $2S$ are equivalent and correspond to the 
pair state on a plane with the relative (REL) motion described by 
angular momentum $m$ and radius $r_m$.
The pair state with the smallest allowed orbit (and largest repulsion) 
has ${\cal R}=1$ on a sphere or $m=1$ on a plane, and larger ${\cal R}$ 
and $m$ means larger average separation.
In the limit of $\lambda/R\rightarrow0$ (i.e., either $2S\rightarrow
\infty$ or $R\rightarrow\infty$), the pair wavefunctions and energies 
calculated on a sphere for ${\cal R}=m$ converge to the planar ones
($\psi_{0m}$ and its energy).

\begin{figure}[t]
\rule{1in}{0in}
\epsfxsize=3.75in
\epsffile{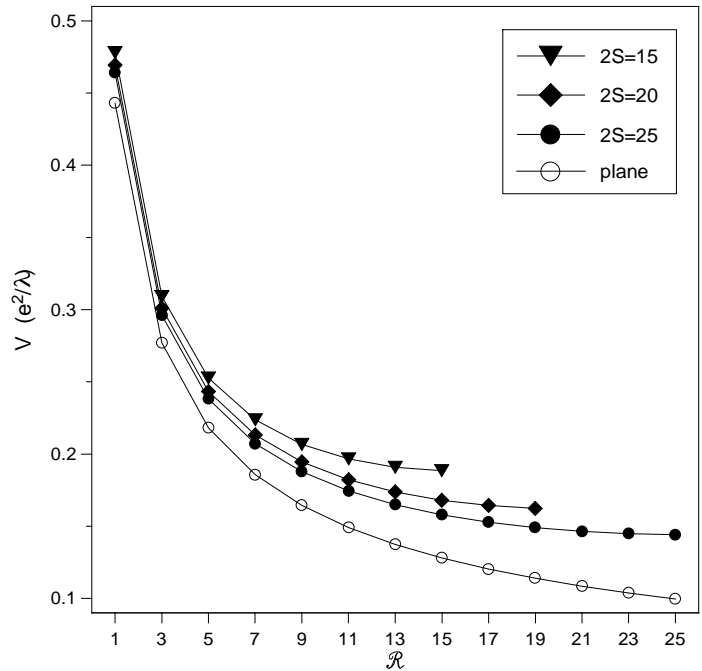}
\caption{
   The pseudopotentials of the Coulomb interaction in the lowest Landau 
   level calculated on a Haldane sphere with $2S=15$, 20, and 25 (solid 
   triangles, diamonds, and circles, respectively), and on a plane 
   (open circles).}
\label{fig3}
\end{figure}

The pseudopotentials $V({\cal R})$ are plotted in figure~\ref{fig3} for 
a number of values of the monopole strength $2S$.
The open circles mark the pseudopotential calculated on a plane 
(${\cal R}=m$).
At small ${\cal R}$ the pseudopotentials rise very quickly with 
decreasing ${\cal R}$ (i.e. separation).
More importantly, they increase more quickly than linearly as 
a function of $L_{12}(L_{12}+1)$.
The pseudopotentials with this property form a class of so-called
``short range'' repulsive pseudopotentials \cite{wojs1}.
If the repulsive interaction has short range, the low energy many 
body states must, to the extent that it is possible, avoid pair 
states the smallest values of ${\cal R}$ (or $m$) and the maximum 
two body repulsion.

\subsection{
   Fractional Grandparentage}
It is well-known in atomic and nuclear physics that eigenfunction 
of an $N$ Fermion system of total angular momentum $L$ can be 
written as
\begin{equation}
\label{eq11}
   \left|l^N,L\alpha\right>=
   \sum_{L_{12}}\sum_{L'\alpha'}G_{L\alpha,L'\alpha'}(L_{12})
   \left|l^2,L_{12};l^{N-2},L'\alpha';L\right>.
\end{equation}
Here, the totally antisymmetric state $\left|l^N,L\alpha\right>$
is expanded in the basis of states $\left|l^2,L_{12};l^{N-2},
L'\alpha';L\right>$ which are antisymmetric under permutation of 
particles 1 and 2 (which are in the pair eigenstate of angular 
momentum $L_{12}$) and under permutation of particles 3, 4, \dots, 
$N$ (which are in the $N-2$ particle eigenstate of angular momentum 
$L'$).
The labels $\alpha$ (and $\alpha'$) distinguish independent states 
with the same angular momentum $L$ (and $L'$).
The expansion coefficient $G_{L\alpha,L'\alpha'}(L_{12})$ is called 
the coefficient of fractional grandparentage (CFGP).

For a simple three Fermion system, equation~(\ref{eq11}) reduces to
\begin{equation}
   \left|l^3,L\alpha\right>=
   \sum_{L_{12}}F_{L\alpha}(L_{12})\left|l^2,L_{12};l;L\right>,
\end{equation}
and $F_{L\alpha}(L_{12})$ is called the coefficient of fractional 
parentage (CFP).
In the lowest Landau level, the individual Fermion angular momentum 
$l$ is equal to $S$, half the monopole strength, and the number of 
independent multiplets of angular momentum $L$ that can be formed by 
addition of the angular momenta of three identical Fermions is given 
in table~\ref{tab3}
\begin{table}
\caption{
   The number of times an $L$ multiplet appears for a system of three
   electrons of angular momentum $l$.
   Top: even values of $2l$; bottom: odd values of $2l$.
   Blank spaces are equivalent to zeros.}
\begin{indented}
\item[]
\begin{tabular}{@{}rrccccccccccccc}
\br
    $2l$&2S=0&2&4&6&8&10&12&14&16&18&20&22&24&26\\
\mr
    2&1&&&&&&&&&&&&&\\
    4&&1&&1&&&&&&&&&&\\
    6&\underline{1}&&1&1&1&&1&&&&&&&\\
    8&&\underline{1}&&2&1&1&1&1&&1&&&&\\
   10&\underline{\underline{1}}&&\underline{1}&\underline{1}&2&1&2&1
      &1&1&1&&1&\\
   12&&\underline{\underline{1}}&&\underline{2}&\underline{1}&2&2&2
      &1&2&1&1&1&1\\
   14&\underline{\underline{\underline{1}}}&&\underline{\underline{1}}
      &\underline{\underline{1}}&\underline{2}&\underline{1}
      &3&2&2&2&2&1&2&1\\
\br
\end{tabular}
\item[]
\begin{tabular}{@{}rrccccccccccccc}
\br
    $2l$&2S=1&3&5&7&9&11&13&15&17&19&21&23&25&27\\
\mr
    3&&1&&&&&&&&&&&&\\
    5&&1&1&&1&&&&&&&&&\\
    7&&\underline{1}&1&1&1&1&&1&&&&&&\\
    9&&\underline{1}&\underline{1}&1&2&1&1&1&1&&1&&&\\
   11&&\underline{\underline{1}}&\underline{1}&\underline{1}&2&2&1&2&1
      &1&1&1&&1\\
   13&&\underline{\underline{1}}&\underline{\underline{1}}&\underline{1}
      &\underline{2}&2&2&2&2&1&2&1&1&1\\
\br
\end{tabular}
\end{indented}
\label{tab3}
\end{table}

Low energy many body states must, to the extent it is possible, avoid 
parentage from pair states with the largest repulsion (pair states with 
maximum angular momenta $L_{ij}$ or minimum ${\cal R}$).
In particular, we expect that the lowest energy multiplets will avoid 
parentage from the pair state with ${\cal R}=1$.
If ${\cal R}=1$, i.e. $L_{12}=2l-1$, the smallest possible value of the 
total angular momentum $L$ of the three Fermion system is obtained by
addition of vectors $\bi{L}_{12}$ (of length $2l-1$) and $\bi{l}_3$ 
(of length $l$), and it is equal to $|(2l-1)-l|=l-1$.
Therefore, the three particle states with $L<l-1$ must not have parentage 
from ${\cal R}=1$.
It is straightforward to show that if $L<l-(2p-1)$, where $p=1$, 2, 3,
\dots, the three electron multiplet at $L$ has no fractional parentage 
from ${\cal R}\le2p-1$.
The multiplets that must avoid one, two, or three smallest values 
of ${\cal R}$ are underlined with an appropriate number of lines in
table~\ref{tab3} and listed in table~\ref{tab4}.
This gives the results in table~\ref{tab4}, the values of $2L$ that 
avoid ${\cal R}=1$, 3, and 5 for various values of $2l$.
\begin{table}
\caption{
   The allowed values of $2L$ for a three electron system that must 
   have ${\cal R}\ge3$, ${\cal R}\ge5$, and ${\cal R}\ge7$.
   The listed $2L$ values correspond to the underlined $L$ multiplets 
   in table~\protect\ref{tab2}.}
\begin{indented}
\item[]
\begin{tabular}{@{}rlll}
\br
  $2l$ & 
  $2L\;({\cal R}\ge3)$ & 
  $2L\;({\cal R}\ge5)$ &
  $2L\;({\cal R}\ge7)$ \\
\mr
  6 & 0                  &         &   \\
  7 & 3                  &         &   \\
  8 & 2                  &         &   \\
  9 & 3, 5               &         &   \\
 10 & 0, 4, 6            & 0       &   \\
 11 & 3, 5, 7            & 3       &   \\
 12 & 2, $6^2$, 8        & 2       &   \\
 13 & 3, 5, 7, $9^2$     & 3, 5    &   \\
 14 & 0, 4, 6, $8^2$, 10 & 0, 4, 6 & 0 \\
\br
\end{tabular}
\end{indented}
\label{tab4}
\end{table}
The $L=0$ states that appear at $2l=6$ (${\cal R}\ge3$), $2l=10$ 
(${\cal R}\ge5$), and $2l=14$ (${\cal R}\ge7$) are the only states 
for these values of $2l$ that can avoid one, two, or three largest
pseudopotential parameters, respectively, and therefore are the 
non-degenerate ($L=0$) ground states.
They are the Laughlin $\nu=1/3$, 1/5, and 1/7 states.

If only a single multiplet belongs to an angular momentum subspace,
its form is completely determined by the requirement that it is an 
eigenstate of angular momentum with a given eigenvalue $L$.
The wavefunction and the type of many body correlations do not 
depend on the form of the interaction pseudopotential.
For interactions that do not have short range, the state that avoids 
the largest two body repulsion (e.g. the $L=0$ multiplet at $2l=6$) 
might not have the lowest total three body interaction energy and 
be the ground state.
If more than one multiplet belongs to a given angular momentum 
eigenvalue (e.g., two multiplets occur at $L=3$ for $2l=8$), the 
interparticle interaction must be diagonalized in this subspace 
(two-dimensional for $2l=8$ and $L=3$).
Whether the lowest energy eigenstate in this subspace has Laughlin
type correlations, i.e. avoids as much as possible largest two body
repulsion, depends critically on the short range of the interaction
pseudopotential.
For the Coulomb interaction, we find that the Laughlin correlations
occur and, whenever possible, the CFP of the lowest lying multiplets
virtually vanishes (it would vanish exactly for an ``ideal'' short 
range pseudopotential which increases infinitely quickly with 
decreasing ${\cal R}$).
For example, for the lower energy eigenstate at $L=3$ and $2l=8$, 
the CFP for ${\cal R}=1$ is less than $10^{-3}$.
A similar thing occurs at $2S=9$ for $L=9/2$, at $2S=10$ for $L=4$ 
and 6, at $2S=11$ for $L=9/2$, 11/2, and 15/2, at $2S=12$ for $L=5$, 
6, 7, and 9, at $2S=13$ for $L=11/2$, 13/2, 15/2, 17/2, and 21/2, 
and at $2S=14$ for $L=6^2$, 7, 8, 9, 10, and 12.
At $2S=14$ for $L=6$ there are three allowed multiplets.
The diagonalization of the Coulomb interaction gives the lowest state 
that avoids ${\cal R}=1$ (CFP $\sim10^{-7}$) and ${\cal R}=3$ (CFP 
$<10^{-2}$), and the next lowest state that avoids ${\cal R}=1$ 
(CFP $<10^{-5}$) but orthogonality to the lowest state requires that 
it has significant parentage from ${\cal R}=3$ (CFP $\approx0.34$).

One can see that the set of angular momentum multiplets $L$ that 
can be constructed at a given value of $2l$ without parentage from
pair states with ${\cal R}=1$ is identical to the set of all allowed
multiplets $L$ at $2l^*=2l-4$.
For a short range repulsion (e.g. the Coulomb repulsion in the lowest 
Landau level), these multiplets will be (to a good approximation)
the lowest energy eigenstates (the appropriate CFP for the actual 
eigenstates will be very small although not necessarily zero).
More generally, in the lowest Landau level (remember that $l=S$), 
the set of multiplets $L$ that can be constructed at given $2S$ 
without parentage from ${\cal R}\le2p-1$ (i.e. with ${\cal R}\ge2p+1$
for all pairs; $p=1$, 2, \dots) is identical to the set of all 
allowed multiplets $L$ at $2S^*=2S-2p(N-1)$.
The multiplets $L$ forming the lowest Coulomb energy band at a given 
$2S$ are all multiplets allowed at $2S^*$.
But $2S^*=2S-2p(N-1)$ is just the effective magnetic monopole strength 
in the mean field CF picture!
Thus the CF picture with $2p$ attached flux quanta simply picks the 
subset of angular momentum multiplets which have no parentage from 
pair states with ${\cal R}\le2p-1$, and neglects the long range part
of the pseudopotential, $V({\cal R})$ for ${\cal R}\ge2p+1$.

\subsection{
   Definition of the Short Range Pseudopotential}
For systems containing more than three Fermions in an angular momentum 
shell, the simple addition of angular momentum to determine the smallest
possible $L$ that has parentage from pair states with $L_{12}=2l-1$ is of
no help.
Instead, we make use of the following operator identity
\begin{equation}
\label{eq13}
   \hat{L}^2 + N(N-2)\;\hat{l}^2 = \sum_{i<j}\hat{L}_{ij}^2.
\end{equation}
Here $\hat{L}=\sum_i\hat{l}_i$ and $\hat{L}_{ij}=\hat{l}_i+\hat{l}_j$.
The identity is easily proved by writing out the expression for 
$\hat{L}^2$ and for $\sum_{i<j}\hat{L}_{ij}^2$ and eliminating 
$\sum_{i<j}(\hat{l}_i\cdot\hat{l}_j)$ from the pair of equations.
Taking matrix elements of equation~(\ref{eq13}) between states 
$\left|l^N,L\alpha\right>$ described by equation~(\ref{eq11}) gives
\begin{eqnarray}
\label{eq14}
   L(L+1)&+&N(N-2)\,l(l+1)
   =\left<l^N,L\alpha\right|
   \sum_{i<j}\hat{L}_{ij}^2
   \left|l^N,L\alpha\right>
\nonumber\\
   &=&{1\over2}N(N-1)\sum_{L_{12}}{\cal G}_{L\alpha}(L_{12})\;
   L_{12}(L_{12}+1),
\end{eqnarray}
where 
\begin{equation}
   {\cal G}_{L\alpha}(L_{12})=
   \sum_{L'\alpha'}\left|G_{L\alpha,L'\alpha'}(L_{12})\right|^2.
\end{equation}
The coefficients of grandparentage satisfy the relation
\begin{equation}
   \sum_{L_{12}} \sum_{L'\alpha'}
   G_{L\alpha,L'\alpha'}(L_{12})\; G_{L\beta,L'\alpha'}(L_{12})
   =\delta_{\alpha\beta}.
\end{equation}
Of course, the energy of the multiplet $\left|l^N,L\alpha\right>$
is given by
\begin{equation}
   E_\alpha(L)={1\over2}N(N-1)
   \sum_{L_{12}} {\cal G}_{L\alpha}(L_{12}) \, V(L_{12}),
\end{equation}
where $V(L_{12})$ is the electron pseudopotential.

It is important to make the following observations:
\begin{enumerate}
\item 
The expectation value of $\sum_{i<j}\hat{L}_{ij}^2$ in a many body 
state $\left|l^N,L\alpha\right>$ increases as $L(L+1)$, but it is 
totally independent of $\alpha$;
\item 
If the pseudopotential $V_H(L_{12})$ were a linear function of 
$\hat{L}_{12}^2$ (we refer to $V_H$ as the ``harmonic pseudopotential''), 
all many body multiplets with the same value of $L$ would be degenerate;
\item 
The difference $\Delta V(L_{12})=V(L_{12})-V_H(L_{12})$ between the 
actual pseudopotential $V$ and its harmonic part $V_H$ lifts this 
degeneracy.
If $N_L$ many body multiplets of $V_H$ occur at angular momentum $L$, 
the anharmonic term $\Delta V$ in the pseudopotential causes them to 
``repel one another'' and results in a band of $N_L$ non-degenerate 
multiplets.
\end{enumerate}
Because the expectation value of $\sum_{i<j}\hat{L}_{ij}^2$ in 
a many body state of angular momentum $L$ increases as $L(L+1)$, 
a strict Hund's rule holds for harmonic pseudopotentials:
For $V_H$ that increases as a function of $L_{12}$, the highest 
energy state is always at the maximum possible value of $L$ equal 
to $L^{\rm MAX}=Nl-N(N-1)/2$, and the lowest energy state is at 
the minimum allowed value of $L$ equal to $L^{\rm MIN}$.
If $V_H$ decreases as a function of $L_{12}$, the opposite occurs:
the lowest energy state is at $L^{\rm MAX}$, and the highest energy 
state is at $L^{\rm MIN}$ (this is a standard Hund's rule of atomic 
physics).

Neither of these Hund's rules may remain true in the presence of 
a large anharmonic term $\Delta V$.
For example, if the number of multiplets $N_L$ at a value slightly 
larger than $L^{\rm MIN}$ is very large compared to $N_{L^{\rm MIN}}$, 
the strong level repulsion due to $\Delta V$ within this $L$ subspace 
can overcome the difference in the expectation values of $V_H$, and 
the lowest eigenvalue of $V$ at $L$ can be lower than that at 
$L^{\rm MIN}$.
However, only very few multiplets occur at large values of $L$:
$N_{L^{\rm MAX}}=1$ (for $M=L=L^{\rm MAX}$, the only state is
$\left|l,l-1,\dots,l-N+1\right>$), $N_{L^{\rm MAX}-1}=0$, 
$N_{L^{\rm MAX}-2}\le1$, $N_{L^{\rm MAX}-3}\le1$, etc.
As a result, breaking of the Hund's rule that refers to the behavior 
of energy at large $L$ requires stronger anharmonicity than at small 
$L$.
For the Coulomb pseudopotential in the lowest Landau level we always 
find that the highest energy indeed occurs at $L^{\rm MAX}$.
However, the ability to avoid parentage from pair states having large 
$L_{ij}$ often favors many body states at small $L>L^{\rm MIN}$ with 
large $N_L$, as prescribed by the CF picture.

The anharmonicity of the Coulomb pseudopotential in the lowest Landau 
level (which increases with increasing $L_{12}$) is critical for the 
behavior of the FQH systems.
We have found that the condition for the occurrence of subbands 
separated by gaps in the energy spectrum, and, in particular, for
the occurrence of non-degenerate incompressible fluid ground states
at specific values of the filling factor, is that the anharmonic 
term $\Delta V(L_{12})$ is positive and increases with increasing 
$L_{12}$.
In other words, the total pseudopotential $V(L_{12})$ must increase
more quickly than linearly as a function of $L_{12}(L_{12}+1)$.

\subsection{
   Hidden Symmetry of the Short Range Repulsion}
From our numerical studies we have arrived at the following 
conjectures:
\begin{enumerate}
\item
The Hilbert space ${\cal H}_{Nl}$ of $N$ identical Fermions each 
with angular momentum $l$ contains subspaces ${\cal H}_{Nl}^{(p)}$ of 
states that have no parentage from ${\cal R}\le2p-1$.
The subspaces $\tilde{\cal H}_{Nl}^{(p)}={\cal H}_{Nl}^{(p)}\setminus
{\cal H}_{Nl}^{(p+1)}$ can be defined; they hold states without
parentage from ${\cal R}\le2p-1$, but with some parentage from 
${\cal R}=2p+1$.
Then
\begin{equation}
   {\cal H}_{Nl}=\tilde{\cal H}_{Nl}^{(0)}\oplus
                 \tilde{\cal H}_{Nl}^{(1)}\oplus
                 \tilde{\cal H}_{Nl}^{(2)}\oplus\dots.
\end{equation}
\item
For an ``ideal'' short range repulsive pseudopotential $V_{\rm SR}$, 
for which $V_{\rm SR}({\cal R})\gg V_{\rm SR}({\cal R}+2)$, the huge 
difference between energy scales associated with different pair states
results in the following (dynamical) symmetry:
(i) subspaces $\tilde{\cal H}_{Nl}^{(p)}$ are the interaction 
eigensubspaces, 
(ii) $p$ is a good quantum number, 
(iii) energy spectrum splits into bands (larger $p$ corresponds 
to lower energy), and 
(iv) energy gap above the $p$th band scales as $V(2p-2)-V(2p)$.
\item
For a finite short range pseudopotential $V$ (increasing more quickly 
than $V_H$ as a function of $L_{12}$), the above symmetry is only
approximate, but the correlation between energy and parentage from 
highly repulsive pair states persists, and so do the gaps in the 
energy spectrum.
The mixing between neighboring subbands is weak, although the 
structure of energy levels within each subband depends on the 
form of $V(L_{12})$ at ${\cal R}\ge2p+1$.
\item
The set of angular momentum multiplets in subspace ${\cal H}_{Nl}^{(p)}$
is identical to ${\cal H}_{Nl^*}$, where $l^*=l-p(N-1)$.
\end{enumerate}
Although at present we do not have a general analytic proof for the 
last conjecture, we have verified it for various small systems and
have not found one for which it would fail.

The above conjectures can be immediately translated into the planar 
geometry.
The harmonic pseudopotential $V_H(m)$, used to define the class of 
short range pseudopotentials, is that of a repulsive interaction 
potential $V(r)$ which is linear in $r^2$.
Then, 
\begin{equation}
   {\cal H}_{\nu}=\tilde{\cal H}_{\nu}^{(0)}\oplus
                  \tilde{\cal H}_{\nu}^{(1)}\oplus
                  \tilde{\cal H}_{\nu}^{(2)}\oplus\dots,
\end{equation}
where ${\cal H}_{\nu}$ is the Hilbert space of electrons filling 
a fraction $\nu$ of an infinitely degenerate Landau level, and
subspaces $\tilde{\cal H}_{\nu}^{(p)}$ contain states without 
parentage from $m\le2p-1$, but with some parentage from $m=2p+1$.
The (approximate) dynamical symmetry holds for the Coulomb interaction, 
and the low energy band ${\cal H}_{\nu}^{(p)}$ contains the same
angular momentum multiplets as ${\cal H}_{\nu^*}$, with $\nu^*$
defined by the CF prescription in equation~(\ref{eq10}).

\begin{figure}[t]
\rule{1in}{0in}
\epsfxsize=3.75in
\epsffile{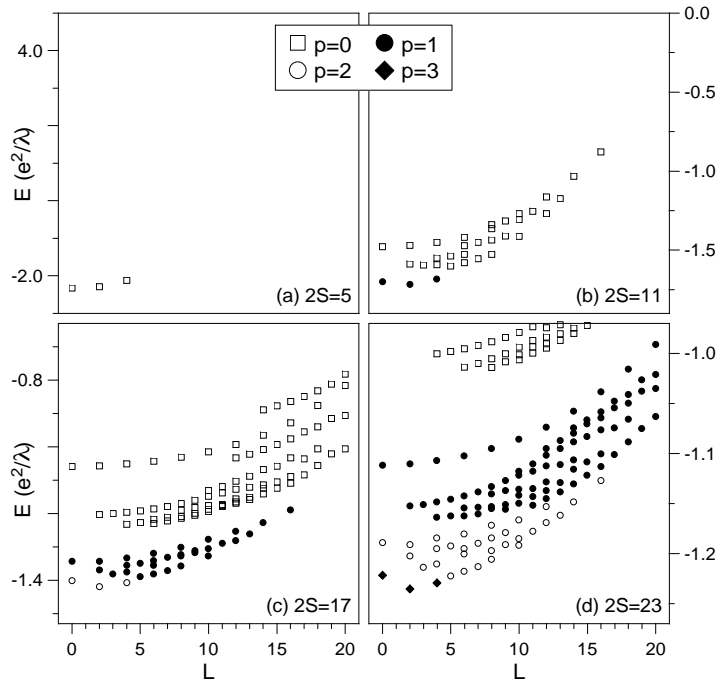}
\caption{
   The energy spectra of four electrons in the lowest Landau level 
   calculated on a Haldane sphere with $2S=5$, 11, 17, and 23.
   All those values of $2S$ are equivalent in the mean field composite 
   Fermion picture (the Chern--Simons transformation with $p=0$, 1, 2, 
   and 3, respectively).
   Different symbols mark states with different numbers of avoided
   pair states with highest energy.}
\label{fig4}
\end{figure}

The validity of our conjectures for systems interacting through the 
Coulomb pseudopotential is illustrated in figure~\ref{fig4} for four
electrons in the lowest Landau level at $2S=5$, 11, 17, and 23.
Different symbols mark bands corresponding to (approximate) subspaces 
${\cal H}_{Nl}^{(p)}$ with different $p$.
The same sets of multiplets reoccur for different $2S$ in bands 
related by ${\cal H}_{Nl}^{(p)}\sim{\cal H}_{Nl^*}$.

\subsection{
   Comparison with Atomic Shells: Hund's Rule}
\begin{figure}[t]
\rule{1in}{0in}
\epsfxsize=3.75in
\epsffile{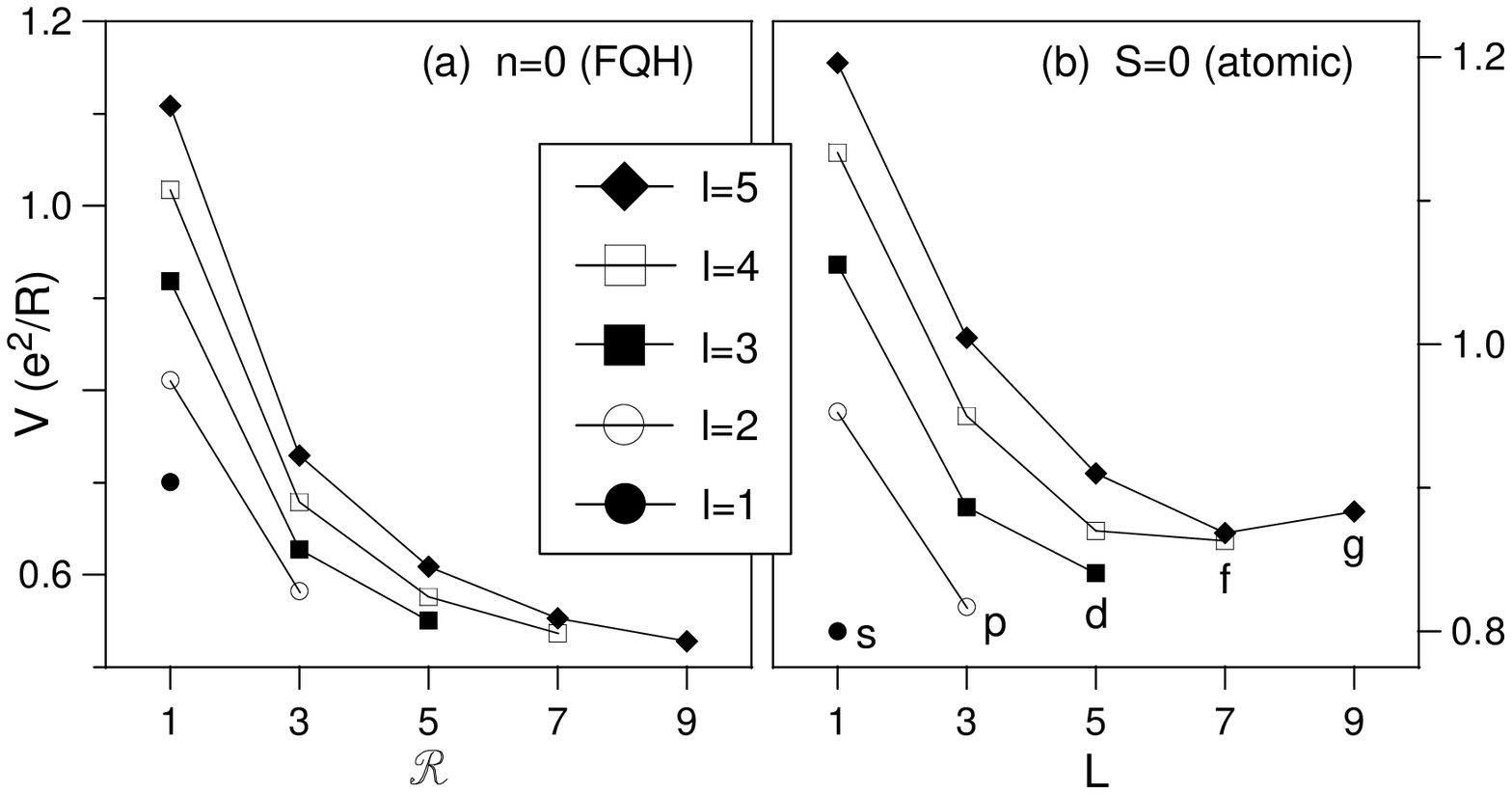}
\caption{
   The pseudopotentials $V$ of the Coulomb interaction for the pair 
   of electrons each with angular momentum $l$:
   (a) lowest Landau level on a Haldane sphere, monopole harmonics, 
   $n=0$, $l=S$, $V$ plotted as a function of relative pair angular 
   momentum ${\cal R}$;
   (b) atomic shell, spherical harmonics, $S=0$, $l=n$,
   $V$ plotted as a function of pair angular momentum $L$.}
\label{fig5}
\end{figure}

\begin{figure}[t]
\rule{1in}{0in}
\epsfxsize=3.75in
\epsffile{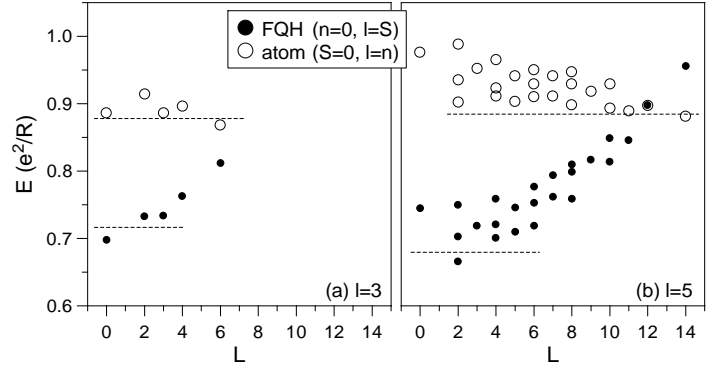}
\caption{
   The energy spectra of four electrons in a degenerate shell of 
   angular momentum $l=3$ (a) and $l=5$ (b), interacting through
   the pseudopotentials of figure~\protect\ref{fig5}:
   open circles -- atomic shell ($S=0$ and $l=n$),
   solid circles -- lowest Landau level ($n=0$ and $l=S$).}
\label{fig6}
\end{figure}

Our conjectures (verified by the numerical experiments) are based on 
the behavior of systems of interacting Fermions partially filling 
a shell of degenerate single particle states of angular momentum $l$.
This is a central problem in atomic physics and in nuclear shell model
studies of energy spectra.
It is interesting to compare the behavior of the spherical harmonics 
of atomic physics with that of the monopole harmonics considered here.
For monopole harmonics $l=S+n$, where $S$ is half of the monopole strength
(and can be integral or half integral) and $n$ is a non-negative integer.
For the lowest angular momentum shell $l=S$.
For spherical harmonics $S=0$ and $l=n$.
If in each case electrons are confined to a 2D spherical surface of 
radius $R$, one can evaluate the pair interaction energy $V$ as 
a function of the pair angular momentum $L_{12}$.
The resulting pseudopotentials, $V({\cal R})$ for the FQH system in the 
lowest Landau level, and $V(L_{12})$ for atomic shells in a zero magnetic 
field, are shown in figure~\ref{fig5} for a few small values of $l$.
In obtaining these results we have restricted ourselves to spin-polarized
shells, so only orbital angular momentum is considered.
It is clear that in the case of spherical harmonics the largest 
pseudopotential coefficient occurs for the lowest pair angular momentum, 
exactly the opposite of what occurs for monopole harmonics.
As a consequence of equation~(\ref{eq13}), which relates the total angular
momentum $L$ to the average pair angular momentum $L_{12}$, the 
standard atomic Hund's rule predicts that the energy of a few electron 
system in an atomic shell will, on the average, decrease as a function 
of total angular momentum, which is opposite to the behavior of energy 
of electrons in the lowest Landau level.
The difference between the energy spectra of electrons interacting
through atomic and FQH pseudopotentials of figure~\ref{fig5} is demonstrated 
in figure~\ref{fig6}, where we plot the result for four electrons in shells
of angular momentum $l=3$ and 5.
The solid circles correspond to monopole harmonics and the open ones 
to spherical harmonics.
Note that at $L^{\rm MAX}$ the former give the highest energy and the 
latter the lowest.
Due to anharmonicity of the pseudopotentials, the behavior of energy
at low $L$ does not always follow a simple Hund's rule for either FQH 
or atomic system.
The FQH ground state for $l=3$ occurs at $L=0$ (this is the $\nu=2/3$ 
incompressible state).
However, for $l=5$, the lowest of the three states at $L=2$ has lower 
energy than the only state $L=0$.
This ground state at $L=2$ contains one quasihole in the Laughlin 
$\nu=1/3$ state and it is the only four electron state at this 
filling in which electrons can avoid parentage from the ${\cal R}=1$ 
pair state.
Exactly opposite happens for the atomic system at $l=5$, where the 
anharmonicity is able to push the highest of the three $L=2$ states 
above the high energy state at $L=0$.

\subsection{
   Higher Landau Levels}
\begin{figure}[t]
\rule{1in}{0in}
\epsfxsize=3.75in
\epsffile{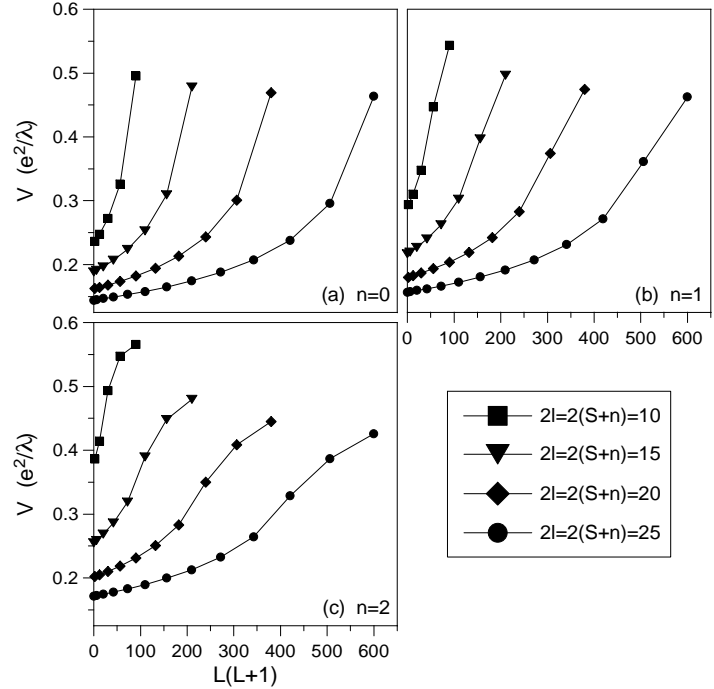}
\caption{
   The pseudopotentials $V$ of the Coulomb interaction in the lowest 
   (a), first excited (b), and second excited (c) Landau levels, 
   calculated on a Haldane sphere, as a function of squared pair 
   angular momentum $L(L+1)$.
   Different symbols correspond to different Landau level degeneracy 
   $2l+1$.}
\label{fig7}
\end{figure}

Thus far we have considered only the lowest angular momentum shell 
(lowest Landau level) with $l=S$
The interaction of a pair of electrons in the $n$th excited shell 
of angular momentum $l=S+n$ can easily be evaluated to obtain the 
pseudopotentials $V(L_{12})$ shown in figure~\ref{fig7}.
Here we compare $V_n(L_{12})$ as a function of $L_{12}(L_{12}+1)$ for 
$n=0$, 1, and 2.
It can readily be observed that $V_{n=0}$ increases more quickly than 
$L_{12}(L_{12}+1)$ in entire range of $L_{12}$, while $V_{n=1}$ and 
$V_{n=2}$ do so only up to certain value of $L_{12}$ (i.e., above 
certain value of ${\cal R}=2l-L_{12}$)
For $n=1$, the $V_{n=1}$ has short range for ${\cal R}\ge3$ but 
is essentially linear in $L_{12}(L_{12}+1)$ from ${\cal R}=1$ to 5.
For $n=2$, the $V_{n=2}$ has short range for ${\cal R}\ge5$ but 
is sublinear in $L_{12}(L_{12}+1)$ from ${\cal R}=1$ to 7.
More generally, we find that the pseudopotential in the $n$th excited 
shell (Landau level) has short range for ${\cal R}\ge2n+1$.

Because the conclusions of the CF picture depend so critically on the
short range of the pseudopotential, they are not expected to be valid 
for all fractional fillings of higher Landau levels.
For example, the ground state at $\nu=2+1/3=7/3$ does not have Laughlin 
type correlations (i.e. electrons in the $n=1$ Landau level do not avoid 
parentage from ${\cal R}=1$) even if it is non-degenerate ($L=0$) and
incompressible (as found experimentally \cite{willet}).

\section{
   Fermi Liquid Model of Composite Fermions}
\label{sec7}
The numerical results of the type shown in figure~\ref{fig1} have been
understood in a very simple way using Jain's composite Fermion picture.
For the ten particle system, the Laughlin $\nu=1/3$ incompressible 
ground state at $L=0$ occurs for $2S=3(N-1)=27$.
The low lying excited states consist of a single QP pair, with the QE 
and QH having angular momenta $l_{\rm QE}=11/2$ and $l_{\rm QH}=9/2$.
In the mean field CF picture, these states should form a degenerate 
band of states with angular momentum $L=1$, 2, \dots, 10.
More generally, $l_{\rm QE}=(N+1)/2$ and $l_{\rm QH}=(N-1)/2$ for the
Laughlin state of an $N$ electron system, and the maximum value of $L$ 
is $N$.
The energy of this band would be $E=\hbar\omega_c^*=\hbar\omega_c/3$, 
the effective CF cyclotron energy needed to excite one CF from the 
(completely filled) lowest to the (completely empty) first excited 
CF Landau level.
From the numerical results, two shortcomings of the mean field CF 
picture are apparent.
First, due to the QE--QH interaction (neglected in the CF picture), 
the energy of the QE--QH band depends on $L$, and the ``magnetoroton'' 
QE--QH dispersion has a minimum at $L=5$.
Second, the state at $L=1$ either does not appear, or is part of the 
continuum (in an infinite system) of higher energy states above the 
magnetoroton band.

At $2S=27-1=26$ and $2S=27+1=28$, the ground state contains a single
quasiparticle (QE or QH, respectively), whose angular momenta 
$l_{\rm QE}=l_{\rm QH}=N/2=5$ result from the CS transformation which 
gives $2S^*=2S-2(N-1)=8$ for QE and 10 for QH (and $l_{\rm QE}=S^*+1$ 
and $l_{\rm QH}=S^*$).
States containing two identical QP's form lowest energy bands at $2S=25$ 
(two QE's) and $2S=29$ (two QH's).
The allowed angular momenta of two identical CF QP's (which are 
Fermions) each with angular momentum $l_{\rm QP}$ are $L=2l_{\rm QP}
-j$ where $j$ is an odd integer.
Of course, $l_{\rm QP}$ depends on $2S$ in the CF picture, and at 
$2S=25$ we have $l_{\rm QE}=S^*+1=S-(N-1)+1=9/2$ yielding $L=0$, 2, 
4, 6, and 8, while at $2S=29$ we have $l_{\rm QH}=S^*=S-(N-1)=11/2$ 
and $L=0$, 2, 4, 6, 8, 10.
More generally, $l_{\rm QE}=(N-1)/2$ and $l_{\rm QH}=(N+1)/2$ in the 
2QE and 2QH states of an $N$ electron system, and the maximum values 
of $L$ are $N-2$ for QE's and $N$ for QH's.
As for the magnetoroton band at $2S=27$, the CF picture does not
account for QP interactions and incorrectly predicts the degeneracy 
of the bands of 2QP states at $2S=25$ and 27.

The energy spectra of states containing more than one CF quasiparticle
can be described in the following phenomenological Fermi liquid picture.
The creation of an elementary excitation, QE or QH, in a Laughlin 
incompressible ground state requires a finite energy, 
$\varepsilon_{\rm QE}$ or $\varepsilon_{\rm QH}$, respectively.
In a state containing more than one Laughlin quasiparticle, QE's and/or 
QH's interact with one another through appropriate QE--QE, QH--QH,
and QE--QH pseudopotentials.

An estimate of the QP energies can be obtained by comparing the energy 
of a single QE (for the $N=10$ electron system, the energy of the ground 
state at $L=N/2=5$ for $2S=27-1=26$) or a single QH ($L=N/2=5$ ground 
state at $2S=27+1=28$) with the Laughlin $L=0$ ground state at $2S=27$.
There can be finite size effects here, because the QP states occur at
different values of $2S$ than the ground state, but using the correct
magnetic length $\lambda=R/\sqrt{S}$ ($R$ is the radius of the sphere) 
in the unit of energy $e^2/\lambda$ at each value of $2S$, and 
extrapolating the results as a function of $N^{-1}$ to an infinite 
system should give reliable estimates of $\varepsilon_{\rm QE}$ and 
$\varepsilon_{\rm QH}$ for a macroscopic system.

The QP pseudopotentials $V_{\rm QP-QP}$ can be obtained by subtracting 
from the energies of the 2QP states obtained numerically at $2S=25$ (2QE), 
$2S=27$ (QE--QH), and $2S=29$ (2QH), the energy of the Laughlin ground 
state at $2S=27$ and two energies of appropriate non-interacting QP's.
As for the single QP, the energies calculated at different $2S$ must be 
taken in correct units of $e^2/\lambda=\sqrt{S}e^2/R$ to avoid finite 
size effects.
This procedure was carried out in references \cite{sitko2,wojs3}.

\begin{figure}[t]
\rule{1in}{0in}
\epsfxsize=3.75in
\epsffile{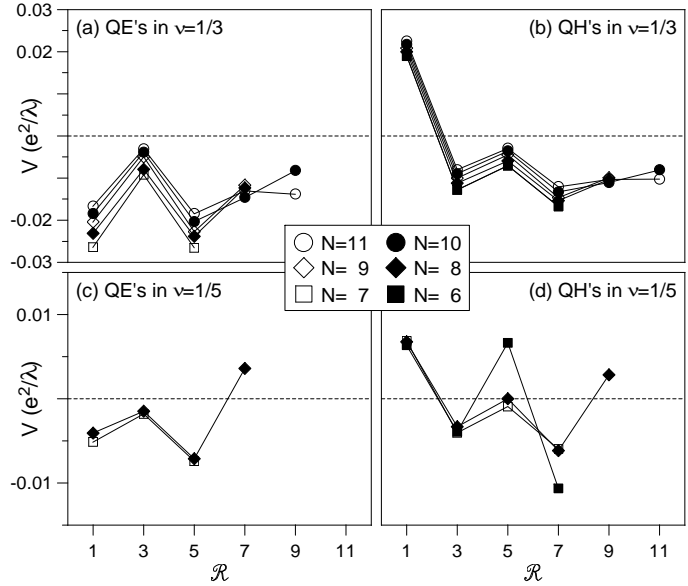}
\caption{
   The pseudopotentials of a pair of quasielectrons (left) and 
   quasiholes (right) in Laughlin $\nu=1/3$ (top) and $\nu=1/5$ 
   (bottom) states, as a function of relative pair angular 
   momentum ${\cal R}$.
   Different symbols mark data obtained in the diagonalization 
   of between six and eleven electrons.}
\label{fig8}
\end{figure}

In figure~\ref{fig8} we plot the QE--QE and QH--QH pseudopotentials for 
Laughlin $\nu=1/3$ and 1/5 states.
As we have seen for two electrons (see figure~\ref{fig3}), the angular 
momentum $L_{12}$ of a pair of identical Fermions in an angular momentum 
shell (or a Landau level) is quantized, and the convenient quantum number
to label the pair states is ${\cal R}=2l_{\rm QP}-L_{12}$ (on a sphere) 
or relative (REL) angular momentum $m$ (on a plane).
When plotted as a function of ${\cal R}$, the pseudopotentials calculated 
for systems containing between six to eleven electrons (and thus for
different QP angular momenta $l_{\rm QP}$) behave similarly and, for 
$N\rightarrow\infty$ (i.e., $2S\rightarrow\infty$), they seem to converge 
to the limiting pseudopotentials $V_{\rm QP-QP}({\cal R}=m)$ describing 
an infinite planar system.

\begin{figure}[t]
\rule{1in}{0in}
\epsfxsize=3.75in
\epsffile{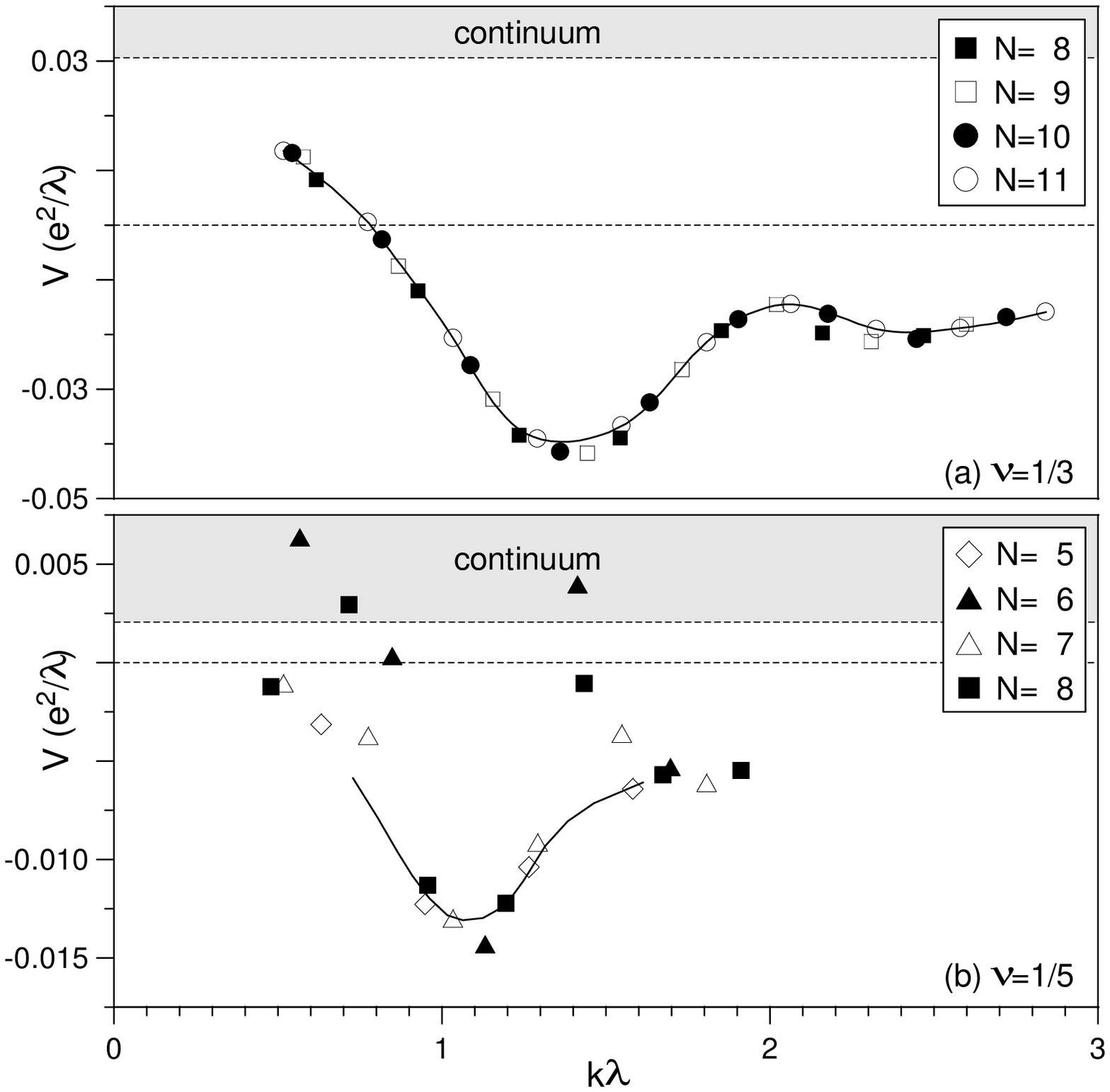}
\caption{
   The pseudopotentials of a quasielectron--quasihole pair in Laughlin 
   $\nu=1/3$ (a) and $\nu=1/5$ (b) states as a function of wavevector 
   $k$.
   Different symbols mark data obtained in the diagonalization 
   of between five and eleven electrons.}
\label{fig9}
\end{figure}

In figure~\ref{fig9} we plot the QE--QH pseudopotentials for Laughlin 
$\nu=1/3$ and 1/5 states.
As for a conduction electron and a valence hole pair in a semiconductor 
(an exciton), the motion of a QE--QH pair which does not carry a net 
electric charge is not quantized in a magnetic field.
The appropriate quantum number to label the states is the continuous
wavevector $k$ (or momentum), which on a sphere is given by $k=L/R=
L/\sqrt{S}\lambda$ (remember that $L$ is given in units of $\hbar$).
When plotted as a function of $k$, the pseudopotentials calculated 
for systems containing between six to eleven electrons fall on 
the same curve that describes a continuous magnetoroton dispersion 
$V_{\rm QE-QH}(k)$ of an infinite planar system (the lines in 
figure~\ref{fig9} are only to guide the eye).
Only the energies for $L\ge2$ are shown in figure~\ref{fig9}, since
the single QE--QH pair state at $L=1$ is either disallowed (hard core) 
or falls into the continuum of states above the magnetoroton band.
The magnetoroton minima for the Laughlin $\nu=1/3$ and 1/5 states
occur at about $k_0=1.4$~$\lambda^{-1}$ and $k_0=1.1$~$\lambda^{-1}$, 
respectively.
The magnetoroton band at $\nu=1/3$ is well decoupled from the continuum 
of higher states because the band width $\sim0.05e^2/\lambda$ is much 
smaller than the energy gap $\varepsilon_{\rm QE}+\varepsilon_{\rm QH}
=0.1e^2/\lambda$ for additional QE--QH pair excitations.
At $\nu=1/5$, the band width $\sim0.015e^2/\lambda$ is closer to the 
single particle gap $\varepsilon_{\rm QE}+\varepsilon_{\rm QH}=0.021
e^2/\lambda$ and the state of two magnetorotons each with $k\approx 
k_0$ can couple to the highest energy QE--QH pair states at $k\le2k_0$.

\begin{figure}[t]
\rule{1in}{0in}
\epsfxsize=3.75in
\epsffile{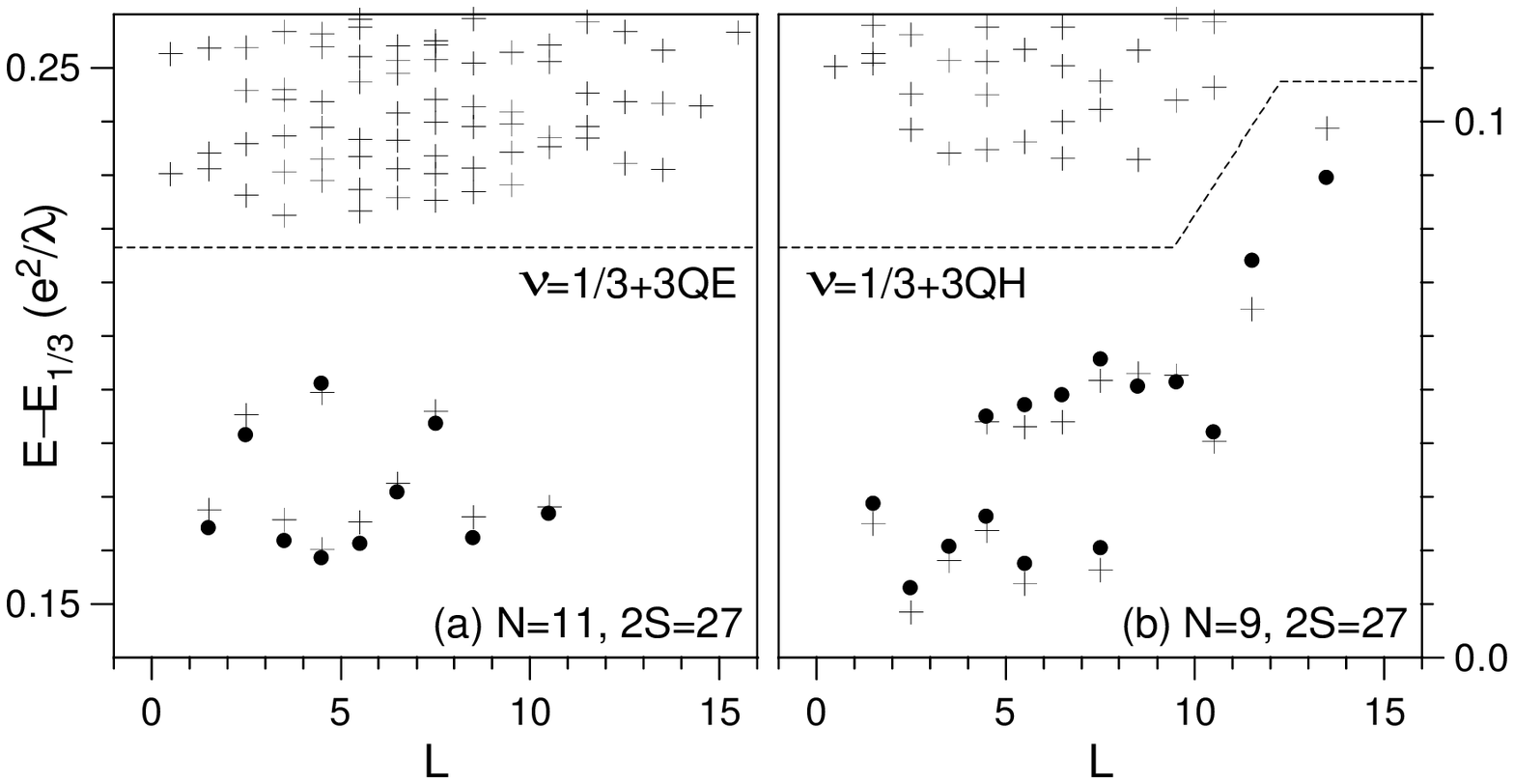}
\caption{
   The energy spectra of three quasielectrons (a) and three
   quasiholes in the Laughlin $\nu=1/3$ state.
   The crosses correspond to the Fermi-liquid calculation using 
   pseudopotentials from figure~\protect\ref{fig8}(a,b); the solid
   circles give exact spectra obtained in full diagonalization of 
   the Coulomb interaction of eleven (a) and nine (b) electrons.}
\label{fig10}
\end{figure}

Knowing the QP--QP pseudopotentials and the bare QP energies allows 
us to evaluate the energies of states containing three or more QP's.
Typical results are shown in figure~\ref{fig10}.
In frame (a) we show the energy spectra of three QE's in the Laughlin
$\nu=1/3$ state of eleven electrons.
The spectrum in frame (b) gives energies of three QH's in the nine 
electron system at the same filling.
The exact numerical results obtained in diagonalization of the eleven
and nine electron systems are represented by plus signs and the Fermi 
liquid picture results are marked by solid circles.
The exact energies above the dashed lines correspond to higher energy 
states that contain additional QE--QH pairs.
It should be noted that in the mean field CF picture which neglects
the QP--QP interactions, all of the 3QP states would be degenerate 
and the energy gap separating the 3QP states from higher states would
be equal to $\hbar\omega_c^*=\hbar\omega_c/3$.
Although the fit in figure~\ref{fig10} is not perfect, it is quite good 
and justifies the use of the Fermi liquid picture to describe 
(compressible) states at $\nu\ne(2p+1)^{-1}$.

\section{
   Composite Fermion Hierarchy}
\label{sec8}
The sequence of Laughlin--Jain states with filling factor $\nu$ given 
by $\nu=\nu^*(1+2p\nu^*)^{-1}$ where $p=1$, 2, \dots, and the CF filling
factor $\nu^*$ is any non-zero integer, is the most prominent set of
condensed states observed experimentally.
However, this sequence (together with the conjugate ``hole'' states,
$\nu\rightarrow1-\nu$) does not contain all odd denominator fractions
the way the Haldane hierarchy scheme does.
The question arises quite naturally of how to treat the CF values of 
$\nu^*$ which are not integers.
The answer leads to the CF approach to the hierarchy of incompressible
quantum fluid ground states \cite{sitko1}.

Consider a state of $N_0$ electrons at a monopole strength $2S_0$ with 
a filling factor $\nu_0$.
The CS transformation that attaches to each electron $2p_0$ flux quanta 
oriented opposite to the applied magnetic field results in the CF system 
at an effective filling factor $\nu_0^*$ given by $(\nu_0^*)^{-1}=
\nu_0^{-1}-2p_0$ and an effective monopole strength $2S_0^*=2S_0-2p_0
(N_0-1)$.
The procedure for handling non-integral values of CF filling factor 
$\nu_0^*$ is to set it equal to $\nu_0^*=n_1+\nu_1$, where $n_1$ is an 
integer and $\nu_1$ is the fractional filling of the CF quasiparticle 
level (same sign as $n_1$ for QE's and opposite for QH's).
Our problem is then that of placing $N_1$ quasiparticles into $2l_1+1$ 
available states of a CF shell (Landau level) of angular momentum $l_1$:
the QE's into the lowest empty shell with $l_1=|S_0^*|+n_1+1$,
or the QH's into the highest filled shell with $l_1=|S_0^*|+n_1$,
We now ignore all completely filled and completely empty CF shells, and 
reapply the CS transformation by setting $S_1=l_1$ and attaching $2p_1$ 
flux quanta to each of the $N_1$ quasiparticles in the partially filled 
CF shell.
This produces a new type of QP's and a new QP filling factor $\nu_1^*$
given by $(\nu_1^*)^{-1}=\nu_1^{-1}-2p_1$.
If $\nu_1^*$ is an integer, we obtain a daughter states in the hierarchy.
If it is not, we write $\nu_1^*=n_2+\nu_2$, where $\nu_2$ represents the
partial filling of the new QP shell, and repeat the mean field CF 
procedure.
This leads to the set of equations:
\begin{equation}
\label{eq17}
   \nu_l^{-1}=2p_l+(n_{l+1}+\nu_{l+1})^{-1},
\end{equation}
where $\nu_l$ is the QP filling factor and $2p_l$ is the number of flux 
quanta attached to each Fermion at the $l$th level of the CF hierarchy.

As an example, consider a system of $N_0=12$ electrons at $2S_0=30$.
We apply the mean field CF approximation by attaching to each electron 
$2p_0=2$ flux quanta.
This gives the effective CF monopole strength $2S_0^*=30-2(12-1)=8$.
The lowest CF shell is filled with nine particles, and there are $N_1=3$ 
quasielectrons in the first excited ($n_1=1$) CF shell of angular 
momentum $l_1=5$.
The filling factor at this level of hierarchy is $\nu_0^*=1+\nu_1$.
We now reapply the CF transformation by attaching $2p_1=4$ flux quanta
to each of $N_1=3$ QE's at $2S_1=10$ and obtain $2S_1^*=10-4(3-1)=2$.
The lowest CF shell of $l_1=1$ is now completely filled yielding
$\nu_1^*=1$.
Using equation~(\ref{eq17}) we obtain $\nu_1^{-1}=4+1^{-1}=5$ and 
$\nu_0^{-1}=2+(1+1/5)^{-1}=17/6$.

\begin{figure}[t]
\rule{1in}{0in}
\epsfxsize=3.75in
\epsffile{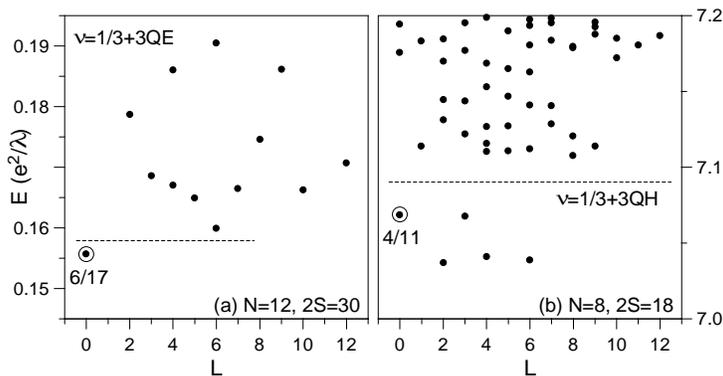}
\caption{
   (a) The low energy spectrum of three quasielectrons in the Laughlin
   $\nu=1/3$ state of twelve electrons calculated using quasielectron
   pseudopotential from figure~\protect\ref{fig8}(a).
   (b) The energy spectrum of three quasiholes in the Laughlin $\nu=1/3$ 
   state of eight electrons obtained in full diagonalization of 
   electron--electron Coulomb interaction.}
\label{fig11}
\end{figure}

If the mean field CF picture worked on all levels of hierarchy, the 
twelve electron system at $2S=30$ should have an incompressible $L=0$ 
ground state corresponding to the filling factor $\nu=6/17$.
In figure~\ref{fig11}(a) we show the low energy sector of the spectrum 
calculated for this system using the Fermi liquid picture
(only the lowest energy states containing 3QE's in the Laughlin
$\nu=1/3$ state are calculated).
Indeed, the $\nu=6/17$ hierarchy ground state at $L=0$ is separated 
from higher states by a small gap in the twelve electron spectrum 
(although it is not clear that this small gap will persist in the 
thermodynamic limit \cite{wojs3}).

Though the CF hierarchy picture seems to work in some cases, there are 
others where it is clearly in complete disagreement with numerical results.
For example, a CF transformation with $2p_0=2$ applied to an $N_0=8$ 
electron system at $2S_0=18$ gives $2S_0^*=18-2(8-1)=4$, $n_1=1$, and 
$N_1=3$ QE's left in the shell with $l_1=3$.
Adding the three QE angular momenta of $l_1=3$ gives a low energy band 
at $L=0$, 2, 3, 4, and 6. 
Reapplication of the CF transformation with $2p_1=2$ gives $2S_1^*=
6-2(3-1)=2$, i.e. the completely filled lowest shell, $\nu_1^*=1$
($n_2=1$ and $\nu_2=0$).
From equation~(\ref{eq17}) we get $\nu_1=1/3$ and $\nu_0=4/11$.
In figure~\ref{fig11}(b) we show the spectrum obtained by exact 
numerical diagonalization of an eight electron system at $2S=18$.
It is apparent that the set of multiplets at $L=0$, 2, 3, 4, and 6 
form the low energy band.
However the reapplication of the mean field CF transformation to the 
three QE's in the $l_1=3$ shell (which predicts an $L=0$ incompressible 
ground state corresponding to $\nu=4/11$) is definitely wrong.

The reason why the CF hierarchy picture does not always work is not 
difficult to understand.
The electron (Coulomb) pseudopotential in the lowest Landau level 
$V_e({\cal R})$ satisfies the ``short range'' criterion (i.e. it 
increases more quickly with decreasing ${\cal R}$ than the harmonic 
pseudopotential $V_H$) in the entire range of ${\cal R}$, which is 
the reason for the incompressibility of the principal Laughlin 
$\nu=(2p+1)^{-1}$ states.
However, this does not generally hold for the QP pseudopotentials
on higher levels of the hierarchy.
In figure~\ref{fig8} we plotted $V_{\rm QE-QE}({\cal R})$ and $V_{\rm 
QH-QH}({\cal R})$ for the $\nu=1/3$ and $\nu=1/5$ Laughlin states of
six to eleven electrons.
Clearly, the QE and QH pseudopotentials are quite different and neither
one decreases monotonically with increasing ${\cal R}$.
On the other hand, the corresponding pseudopotentials in $\nu=1/3$ 
and 1/5 states look similar, only the energy scale is different.
The convergence of energies at small ${\cal R}$ obtained for larger $N$ 
suggests that the maxima at ${\cal R}=3$ for QE's and at ${\cal R}=1$ 
and 5 for QH's, as well as the minima at ${\cal R}=1$ and 5 for QE's 
and at ${\cal R}=3$ and 7 for QH's, persist in the limit of large $N$ 
(i.e. for an infinite system on a plane).
Consequently, the only incompressible daughter states of Laughlin 
$\nu=1/3$ and 1/5 states are those with $\nu_{\rm QE}=1$ or $\nu_{\rm 
QH}=1/3$ and (maybe) $\nu_{\rm QE}=1/5$ and $\nu_{\rm QH}=1/7$.
It is clear that no incompressible daughter states of the parent 
Laughlin $\nu=1/3$ state will form at e.g. $\nu=4/11$ ($\nu_{\rm QE}=
1/3$) or 4/13 ($\nu_{\rm QH}=1/5$), but that they will form (at least, 
in finite systems \cite{wojs3}) at $\nu=6/17$ ($\nu_{\rm QE}=1/5$)
or 4/13 ($\nu_{\rm QH}=1/7$).

From the CF hierarchy scheme we find the Jain--Laughlin states when 
the CS transformation is applied directly to electrons (or to holes 
in a more than half-filled level).
These states occur at integral values of $\nu^*$, the effective CF
filling factor, and correspond to completely filling a QP shell.
For example, the $\nu=2/5$ state occurs when $\nu^*=2$, and the CF's
in the first excited shell (which are Laughlin QE's of the $\nu=1/3$
state) have $\nu_{\rm QP}=1$.
The angular momenta of the two lowest CF shells are $l_0^*=|S^*|$ and
$l_1^*=|S^*|+1$, so they contain $2l_0^*+1$ and $2l_1^*+1$ states,
respectively.
Since $\nu_{\rm QP}=1$, there are $N_{\rm QP}=2l_1^*+1$ CF 
quasiparticles.
The total number of states filled by the $N$ Fermions is $(2l_0^*+1)
+(2l_1^*+1)=2N_{\rm QP}-2$, giving $N=2N_{\rm QP}-2$.
For an infinite system this is just Haldane's relation between the
number of quasiparticles and the number of electrons, $N=2qN_{\rm 
QP}$, for the integer $q=1$.
This demonstrates that integrally filled CF shells correspond to 
$\nu_{\rm QP}=1$, a completely filled shell of Laughlin QP's.
Adding new Fermions to a system with $\nu_{\rm QP}=1$ requires creating
a new type of QP's, and the counting of available QP states turns out
to be exactly the same in the CF hierarchy and Haldane's Boson hierarchy
picrures.
Integral CF filling (i.e., $\nu_{\rm QP}=1$) gives a valid mean field
picture independent of QP--QP interactions provided that the gap for
creating new QP's is positive.
When $\nu^*$ is non-integral, the mean field picture is valid only 
at values of $L$ for which the ``short range'' requirement on the
pseudopotential $V_{\rm QP-QP}(L)$ is satisfied.
The form of the QP--QP interactions obtained from our numerical 
calculations makes it clear that the mean field approximation is not 
valid at certain quasiparticle fillings (e.g. for $\nu_{\rm QP}=1/3$ 
filling of the quasielectron levels of the electron $\nu=1/3$ state).

\section{
   Systems Containing Electrons and Valence Band Holes}
\label{sec9}
There has been a great deal of interest in photoluminescence (PL) of
2D systems in high magnetic fields.
An important ingredient in understanding PL is the negatively charged 
exciton ($X^-$).
The $X^-$ consists of two electrons bound to a valence band hole.
If the total spin of the pair of electrons, $J_e$, is zero, the $X^-$ 
is said to be a singlet ($X^-_s$); if $J_e=1$ the $X^-$ is called 
a triplet ($X^-_t$).
Only the $X^-_s$ is bound in the absence of a magnetic field, but in
infinite magnetic field (so that only a single Landau level is relevant)
only the $X^-_t$ is bound in a 2D system.
It often occurs that the photoexcited hole is separated from the plane
of the electron system by a small distance (this can happen, e.g., in
wide GaAs quantum wells when the electron gas is confined to one 
GaAs/AlGaAs interface by remote ionized donors, and the photoexcited 
holes reside close to the other GaAs/AlGaAs interface).
Several remarkable effects associated with electron--hole systems and 
charged excitons can be understood using the composite Fermion picture.

\subsection{
   Charged Exciton and the Hidden Symmetry in the Lowest Landau Level}
\begin{figure}[t]
\rule{1in}{0in}
\epsfxsize=3.75in
\epsffile{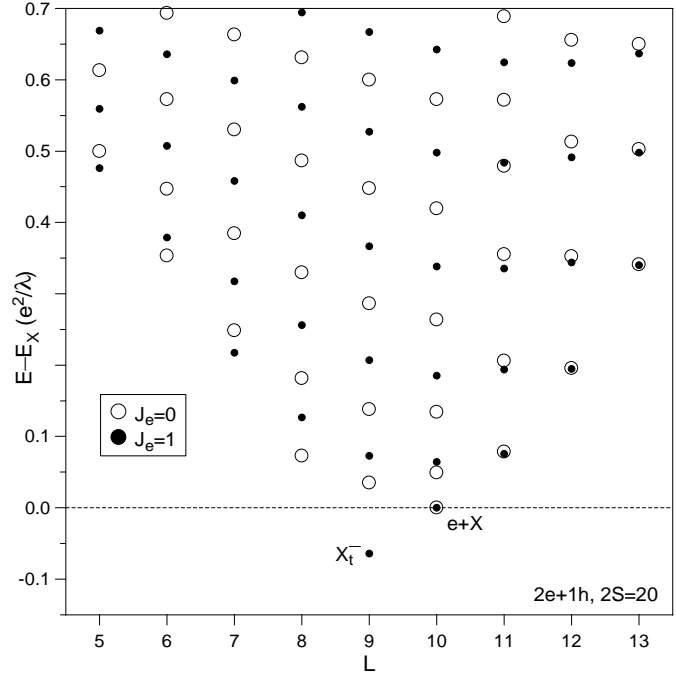}
\caption{
   The energy spectrum (binding energy vs.\ angular momentum) of a 
   two-electron--one-hole system in the lowest Landau level at $2S=20$.
   Open and solid circles mark singlet and triplet spin 
   configurations, respectively.}
\label{fig12}
\end{figure}

First let us consider the idealized 2D system at so large a magnetic 
field that only the lowest electron and hole Landau levels need be 
considered.
The energy spectrum for a two-electron--one-hole system at $2S=20$
is shown in figure~\ref{fig12}.
The triplet $X^-$ with angular momentum $l_{X^-}=S-1$ is the only bound 
state, with binding energy $\sim0.05e^2/\lambda$.
A pair of (unbound) singlet and triplet states occur at the energy equal 
exactly to the exciton energy $E_X$.
In these so-called ``multiplicative'' states a neutral exciton $X$ in 
its ground state is decoupled from the second electron.
Addition of exciton and electron angular momenta $L_X=0$ and $l_e=S$ 
gives a state of total angular momentum $L=S$, and addition of two 
electron spins of $1/2$ gives both $J_e=0$ and 1 spin configurations.

The occurrence of unbound states at $E=E_X$ and $L=S$ is a manifestation 
of the following ``hidden symmetry:''
Because of the exact overlap of electron and hole orbitals in the lowest
Landau level (scaled with the same magnetic length $\lambda$), and thus 
independence of the strength of interaction of the type of particles 
involved, the commutator of an operator $d_X^\dagger$ that creates an 
exciton in its $L_X=0$ ground state (on a sphere, $d_X^\dagger=\sum_m 
(-1)^mc_m^\dagger h_m^\dagger$, where $c_m^\dagger$ and $h_m^\dagger$ 
are electron and hole creation operators), with the interaction 
Hamiltonian $H$ is $[H,d_X^\dagger]=E_Xd_X^\dagger$.
As a result, if $\Psi$ is an eigenstate of $N_e$ electrons and $N_h$ 
holes with an eigenenergy $E$ and angular momentum quantum numbers $L$ 
and $M$, then the multiplicative state $d_X^\dagger\Psi$ of $N_e+1$ 
electrons and $N_h+1$ holes is also an eigenstate with energy $E+E_X$ 
and the same $L$ and $M$.
A good quantum number conserved due to the ``hidden symmetry'' is the 
number of decoupled excitons, $N_X$.
In particular, the ground state for $N_e=N_h=N$ is the totally 
multiplicative state $(d_X^\dagger)^N\left|{\rm vac}\right>$ with
$N_X=N$; for an infinite system this ground state can be viewed as 
a Bose condensate of non-interacting excitons.
It can be readily found that the application of the PL operator that
annihilates an optically active exciton ($d_X$) reduces its $N_X$ by 
one, and therefore that only the multiplicative electron--hole states 
with $N_X>0$ are optically active (have non-vanishing PL intensity).
In figure~\ref{fig12}, the two multiplicative states at $E=E_X$ and 
$L=S$ have $N_X=1$, and all others have $N_X=0$.

It is essential to realize that two independent symmetries forbid the
recombination of a triplet $X^-$ ground state in figure~\ref{fig12}:
\begin{itemize}
\item
Due to the 2D translational/rotational space invariance, the PL 
operator $d_X$ conserves two angular momentum quantum numbers. 
On a sphere, these are is $L$ and $M$, and the resulting optical 
selection rule allows only a state with $L=S$ to decay by $e$--$h$
recombination. 
On a plane, these are the total ($L_{\rm TOT}$) and center-of-mass 
($L_{\rm CM}$) angular momenta and the radiative channel for an $X^-$ 
is that of $L_{\rm REL}\equiv L_{\rm TOT}-L_{\rm CM}=0$.
This (geometrical) symmetry can be broken by collisions, but persists
in systems with a finite quantum well width, finite electron and hole
layer separation, or Landau level mixing.
\item
Due to the equal strength of $e$--$e$, $h$--$h$, and $e$--$h$ 
interactions, $N_X$ is a good quantum number.
Since $N_X$ is decreased in a PL process, only the multiplicative
($N_X>0$) states are radiative.
This (dynamical) symmetry is not broken by collisions, and requires
breaking electron--hole orbital symmetry.
\end{itemize}
Since a number of independent factors are needed to allow for the 
recombination of a triplet $X^-$, this complex (in narrow and symmetric 
quantum wells and in high magnetic fields) is expected to be a well 
defined long-lived quasiparticle.
The correlations, optical properties, etc. are expressed more easily
in terms of this quasiparticle than in terms of individual electrons 
and holes.
The finite angular momentum of an $X^-$ in spherical geometry (decoupling 
of the CM excitations from the REL motion on a plane) can be viewed as 
the formation of a degenerate Landau level of this (charged) quasiparticle.
As will be shown later, the interaction of $X^-$ quasiparticles with
one another and with electrons can be described using the ideas familiar
in the context of FQH systems (Laughlin correlations, composite Fermions, 
parentage, etc.).

\subsection{
   Interaction of Charged Excitons}
\begin{figure}[t]
\rule{1in}{0in}
\epsfxsize=3.75in
\epsffile{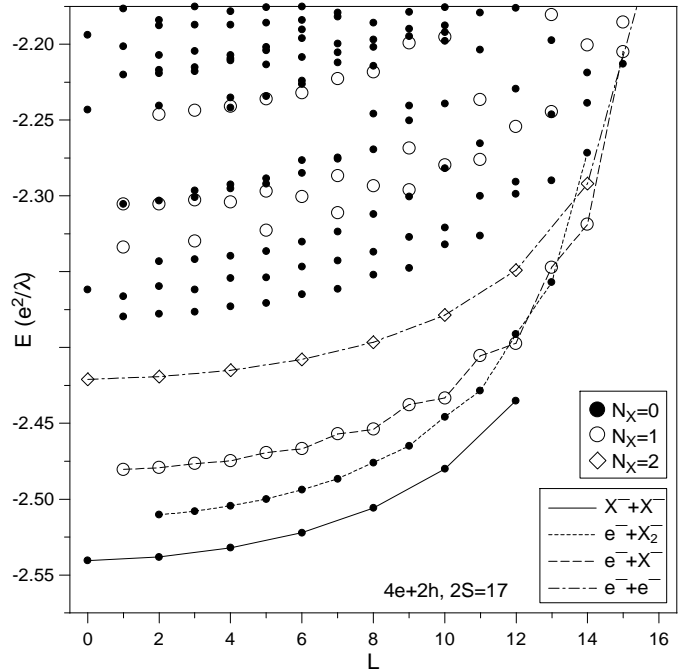}
\caption{
   The energy spectrum of a four-electron--two-hole system in the
   lowest Landau level calculated on a Haldane sphere with $2S=17$.
   Different symbols mark states with zero, one, or two decoupled
   excitons.
   The lines connect states identified as pseudopotentials of different
   pairs of bound charged complexes.}
\label{fig13}
\end{figure}

The simplest system in which to study $X^-$--$X^-$ interaction contains 
four electrons and two holes.
Its energy spectrum at $2S=17$ is shown in figure~\ref{fig13}.
The low energy spectrum is characterized by four bands which we have
identified as follows:
\begin{enumerate}
\item 
The lowest band taking on all even values between $L=0$ and 12
consists of a pair of charged excitons $X^-$ (each with angular 
momentum $l_{X^-}=S-1$);
\item 
The next band contains an electron with $l_e=S$ and a negatively charged 
biexciton $X_2^-$ (a bound state of an $X$ and an $X^-$) with angular
momentum $l_{X_2^-}=S-2$; the allowed $L$ values go from $|l_e-l_{X_2^-}|
=2$ to $l_e+l_{X_2^-}-1=14$;
\item 
A band of multiplicative states containing an $X$, an $X^-$, and 
an electron; it begins at $L=|l_e-l_{X^-}|=1$ and goes to $L=l_e
+l_{X^-}-1=15$;
\item 
A band of multiplicative states containing two neutral excitons and 
two free electrons; it takes on all even values of $L$ between zero 
and $2l_e-1=16$.
\end{enumerate}
One interesting feature of figure~\ref{fig13} is that it gives us the
effective pseudopotential $V_{AB}(L)$ for the interaction of the pair 
of Fermions $AB$ (where $A$ and $B$ can be $e$, $X^-$, $X_2^-$, etc.)
as a function of angular momentum.
As for electrons, it is convenient to use the relative pair angular 
momentum ${\cal R}=l_A+l_B-L$.
For identical Fermions with angular momentum $l$, the allowed values 
of $L$ are $2l-j$, where $j$ is an odd integer, i.e., ${\cal R}=1$,
3, 5, \dots, and ${\cal R}\le2l$.
For distinguishable Fermions $A$ and $B$, all values of $L$ between
$|l_A-l_B|$ and $|l_A+l_B|$ are expected, i.e., ${\cal R}=0$, 1, 2,
\dots, and ${\cal R}\le2\min(l_A,l_B)$. 
However, our numerical results display a ``hard core'' repulsion for
composite particles, and one or more of the pair states with the 
largest values of $L$ (smallest ${\cal R}$) are forbidden (i.e. the
corresponding pseudopotential parameters are effectively infinite).
For $A=X_n^-$ and $B=X_m^-$, the smallest allowed value of ${\cal R}$
is given by 
\begin{equation}
   {\cal R}^{\rm MIN}_{AB}=2\min(n,m)+1.
\end{equation}
The identification of pair states $AB$ in figure~\ref{fig13} (as marked 
with lines) was possible by comparing the displayed $4e$--$2h$ spectrum 
with the pseudopotentials of point charge particles with appropriate 
angular momenta $l_A$ and $l_B$ and binding energies $\varepsilon_A$ 
and $\varepsilon_B$ \cite{wojs2}.
The appropriate values of angular momenta $l_A$ and $l_B$, and of 
the binding energies $\varepsilon_A$ and $\varepsilon_B$ are obtained
by diagonalizing smaller systems (e.g. the $2e$--$1h$ system in 
figure~\ref{fig12} for an $X^-$), and the point charge pseudopotentials
are used to approximate the $AB$ interaction.
The approximate $AB$ energies obtained in this way are rather close 
to the exact $4e$--$2h$ energies.
This implies that, due to different energy scales, the internal dynamics 
of charged excitons is weakly coupled to their scattering off one 
another or off electrons, and allows for the interpretation of an 
electron--hole system in terms of well defined charged excitonic 
quasiparticles interacting with one another and with excess electrons 
through Coulomb like forces.
Slight difference between the actual pseudopotentials in figure~\ref{fig13} 
and the pseudopotentials of point charge particles comes from the larger
size of charged excitons and their (nearly frozen) internal degrees 
of freedom.
The latter can be accounted for phenomenologically by attributing each 
type of composite particles a finite electric polarizability to describe
their induced electric dipole moment in the presence of an electric field 
of other charged particles.
Due to an increased charge isotropy, the polarization effects are expected 
to be greatly reduced in larger systems, and disappear completely in the 
fluid type states discussed in the following paragraphs.

\subsection{
   Generalized Composite Fermion Picture for Charged Excitons}
Suppose we have a system of different (distinguishable) charged Fermions 
($A$, $B$, \dots).
They can be distinguished either because they are different species 
(e.g., electrons and charged excitons) or because they are confined to
different, spatially separated layers.
If all particles in such system repel one another through short range 
pseudopotentials (as defined for the electron FQH systems), one can 
think of many body states with Laughlin-type correlations 
\cite{laughlin2,halperin1} given by a generalized (compare 
equation~(\ref{eq5})) Laughlin--Jastrow factor
\begin{equation}
\label{eqcorrel}
   \prod(z_i^{(A)}-z_j^{(B)})^{m_{AB}},
\end{equation}
where $z_i^{(A)}$ is the complex coordinate for the position of the 
$i$th Fermion of type $A$, and the product is over all pairs.
The restrictions on the integers $m_{AB}$ are that $m_{AA}$ must be 
odd, $m_{BA}=m_{AB}$, and $m_{AB}$ must not be smaller than certain 
minimum values ${\cal R}_{AB}^{\rm MIN}$ to avoid the infinite hard 
cores for all pairs.
In a state with correlations given by equation~(\ref{eqcorrel}), 
a number of pair states with largest repulsion are avoided for each 
pair, ${\cal R}_{AB}\ge m_{AB}$.
This is equivalent to saying that the high energy collisions (in 
which any pair of particles would come very close to one another) 
are forbidden in such state.
This intuitive property of the Laughlin fluid states will be very 
useful in the discussion of collision assisted $X^-$ recombination.

A generalized CF picture can be constructed for a system with Laughlin 
correlations.
In this picture, fictitious flux tubes carrying an integral number 
of flux quanta $\phi_0$ are attached to each particle.
In the multi-component system, each particle of type $A$ carries 
flux $(m_{AA}-1)\phi_0$ that couples only to charges on all other 
particles of the same type $A$, and fluxes $m_{AB}\phi_0$ that couple 
to charges on all particles of other types $B$ ($A$ and $B$ are any 
of the types of Fermions).
On a sphere, the effective monopole strength seen by a CF of type $A$ 
(CF-$A$) is
\begin{equation}
   2S_A^*=2S-\sum_b(m_{AB}-\delta_{AB})(N_B-\delta_{AB}).
\label{eq2}
\end{equation}
For different multi-component systems we expect generalized Laughlin 
incompressible states (for two components denoted as $[m_{AA},m_{BB},
m_{AB}]$) when all the hard core pseudopotentials are avoided and 
CF's of each kind fill completely an integral number of their CF 
shells (e.g. $N_A=2l_A^*+1$ for the lowest shell).
In other cases, the low lying multiplets are expected to contain 
different kinds of CF quasiparticles (generalized QE's or QH's),
QP-$A$, QP-$B$, \dots, in the neighboring incompressible state.
It is interesting to realize that the effective monopole strengths
$2S^*_A$, i.e. the effective magnetic fields $B^*_A$ seen by particles
of different type are not generally equal.
One can think of effective CS charges and fluxes of different colors, 
but the resulting number of different effective CF magnetic fields of 
different color can no longer be regarded as physical reality, and no 
cancellation between gauge and Coulomb interactions is possible.

\begin{figure}[t]
\rule{1in}{0in}
\epsfxsize=3.75in
\epsffile{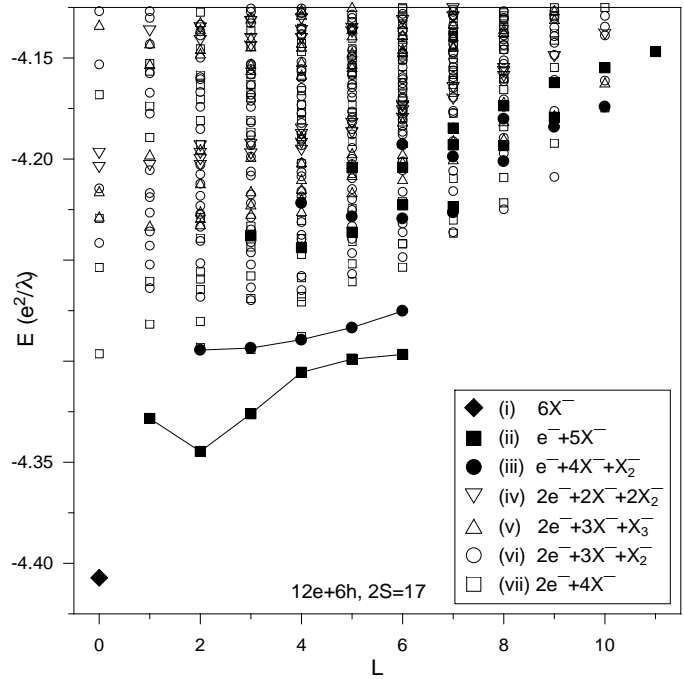}
\caption{
   The approximate lowest energy bands corresponding to different 
   combinations of six bound charged complexes interacting through 
   appropriate pseudopotentials, in the twelve-electron--six-hole 
   spectrum in the lowest Landau level, calculated on a Haldane sphere 
   with $2S=17$.
   The lines mark lowest subbands of two lowest excited bands.}
\label{fig14}
\end{figure}

The multi-component (multi-color) CF picture can be applied to 
electrons and charged excitons in an electron--hole system.
We have checked that the pseudopotentials describing interaction 
of identical composite particles in figure~\ref{fig13} all satisfy 
the short range criterion in the entire range of ${\cal R}$.
For a pair of different particles, the pseudopotential may increase 
sufficiently quickly for some values of ${\cal R}$ but not the 
others and, for example, for $e^-$ and $X^-$ only the correlations 
described by odd exponents $m_{e^-X^-}$ are expected to occur.
As an example, let us consider the $12e$--$6h$ system.
In figure~\ref{fig14} we present its low energy spectrum at $2S=17$,
calculated by diagonalizing systems of different combinations of
electrons and composite particles interacting through effective
pseudopotentials determined in figure~\ref{fig13}.
The following combinations (groupings of $12e$ and $6h$ into bound
complexes) have the highest total binding energy and thus form the 
lowest energy bands in the $12e$--$6h$ spectrum:
(i) $6X^-$, 
(ii) $e^-$--$5X^-$,
(iii) $e^-$--$4X^-$--$X_2^-$,
(iv) $2e^-$--$2X^-$--$2X_2^-$,
(v) $2e^-$--$3X^-$--$X_3^-$,
(vi) $2e^-$--$3X^-$--$X_2^-$,
(vii) $2e^-$--$4X^-$.
Groupings (ii), (vi), and (vii) also contain neutral excitons that
however do not interact with charged particles due to the hidden 
symmetry.
For each of these groupings, the CF transformation can be applied to 
determine correlations and identify number and type of quasiparticles
that occur in the lowest energy states.
For example, for groupings (i)--(iii) the generalized CF picture makes 
the following predictions.
\begin{description}
\item{(i)}
For $m_{X^-X^-}=3$ we obtain the Laughlin $\nu=1/3$ state with total 
angular momentum $L=0$.
Because of the hard core of $V_{X^-X^-}$, this is the only state of
this grouping.
\item{(ii)}
We set $m_{X^-X^-}=3$ and $m_{e^-X^-}=1$, 2, and 3.
For $m_{e^-X^-}=1$ we obtain $L=1$, 2, $3^2$, $4^2$, $5^3$, $6^3$, $7^3$, 
$8^2$, $9^2$, 10, and 11.
For $m_{e^-X^-}=2$ we obtain $L=1$, 2, 3, 4, 5, and 6.
For $m_{e^-X^-}=3$ we obtain $L=1$.
\item{(iii)}
We set $m_{X^-X^-}=3$, $m_{e^-X_2^-}=1$, $m_{X^-X_2^-}=3$, and 
$m_{e^-X^-}=1$, 2, or 3.
For $m_{e^-X^-}=1$ we obtain $L=2$, 3, $4^2$, $5^2$, $6^3$, $7^2$, 
$8^2$, 9, and 10.
For $m_{e^-X^-}=2$ we obtain $L=2$, 3, 4, 5, and 6.
For $m_{e^-X^-}=3$ we obtain $L=2$.
\end{description}
In groupings (ii) and (iii), the sets of multiplets obtained for higher 
values of $m_{e^-X^-}$ are subsets of the sets obtained for lower values, 
and we would expect them to form lower energy bands since they avoid 
additional small values of ${\cal R}_{e^-X^-}$.
However, note that the (ii) and (iii) states predicted for $m_{e^-X^-}=3$ 
(at $L=1$ and 2, respectively) do not form separate bands in 
figure~\ref{fig14}.
This is because $V_{e^-X^-}$ increases more slowly than linearly as 
a function of $L(L+1)$ in the vicinity of ${\cal R}_{e^-X^-}=3$ 
(see figure~\ref{fig13}).
In such case the CF picture fails \cite{wojs1,wojs3}.

Our conclusion is that different kinds of long-lived Fermions 
(electrons and different charged excitonic complexes) formed in 
an electron--hole plasma in high magnetic fields can exhibit 
generalized incompressible FQH ground states with Laughlin-type 
correlations, and that these states can be described using 
a generalized CF model.

\subsection{
   Spatially Separated Electron--Hole System}
Even in very high magnetic fields (in the lowest Landau level), an 
asymmetry between $e$--$e$, $h$--$h$, and $e$--$h$ interactions can 
be introduced by spatially separating 2D electron and hole layers.
Such separation, which occurs for example in asymmetrically doped wide 
quantum wells, breaks the hidden symmetry and allows for a rich
photoluminescence (PL) spectrum, which (unlike that for a co-planar 
system) can be therefore used as a probe of the low lying 
electron--hole states.

\begin{figure}[t]
\rule{1in}{0in}
\epsfxsize=3.75in
\epsffile{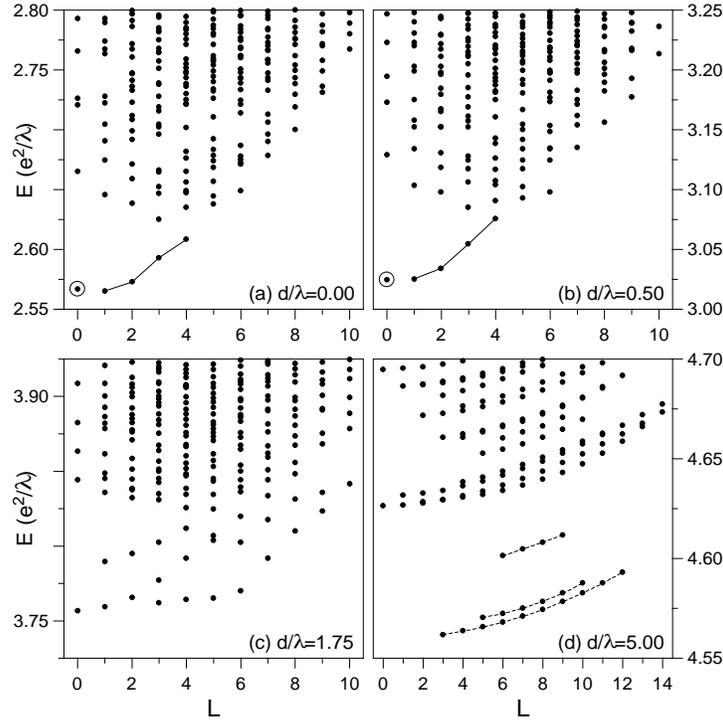}
\caption{
   The energy spectrum of a seven-electron--one-hole system in the 
   lowest Landau level calculated on a Haldane sphere at $2S=15$, 
   for different values of the separation $d$ between electron and 
   hole planes.
   In frames (a) and (b), the circle marks a multiplicative state 
   and solid lines mark states containing a charged exciton $X^-$.
   In frame (d), the dashed lines mark three lowest bands.}
\label{fig15}
\end{figure}

Let us consider an ideal system, in which electrons and holes occupy
2D parallel planes separated by a distance $d$.
The interaction potentials are $V_{ee}(r)=V_{hh}(r)=1/r$ and 
$V_{eh}(r)=-1/\sqrt{r^2+d^2}$.
The energy spectrum of a seven-electron--one-hole system is shown 
in figure~\ref{fig15} for $2S=15$ and values of $d$ going from 0 to 5 
(measured in units of the magnetic length $\lambda$).
For $d=5\lambda$, the $e$--$h$ interaction is weak and, as a first 
approximation, we can say that that the lowest band of states will 
consist of the lowest CF band of the electron system plus the (constant) 
hole energy.
The allowed angular momenta will be given by $\bi{L}_e$, the angular 
momenta of the low lying electron states, added to the hole angular 
momentum $\bi{l}_h$ of length $l_h=S=15/2$.
At $2S=15$, the CF picture for the electrons gives $2S^*=2S-2p(N-1)
=15-2(7-1)=3$.
The seven electrons fill the $l_0^*=3/2$ shell plus three of the QE 
states in the shell $l_{\rm QE}=5/2$.
The resulting electron angular momenta are $L_e=3/2$, 5/2, and 9/2.
This gives three bands of low lying states, with total angular momenta
$6\le L\le 9$, $5\le L\le 10$, and $3\le L\le 12$, respectively.
These three bands can be clearly distinguished at $d=5\lambda$ and 
the states within each band become nearly degenerate at $d\sim10\lambda$.

For $d=0$, it is more useful to consider bound excitonic complexes
($X$ and $X^-$) and Laughlin quasiparticles of the $e^-$--$X^-$ fluid.
First consider the multiplicative state with a single $X$ and six 
electrons. 
At $2S=15$ six electrons have the Laughlin $\nu=1/3$ ground state since 
$2S^*=15-2(6-1)=5$ gives a CF shell which accommodates all six CF's.
This is the lowest state at $L=0$, marked with a circle in frame (a).
For a charge configuration containing one $X^-$ and five electrons,
we can use the generalized CF model with $m_{e^-e^-}=m_{e^-X^-}=2$.
This gives $2S_e^*=2S-m_{e^-e^-}(N_e-1)-m_{e^-X^-}=5$ and $2S_{X^-}^*
=2S-m_{e^-X^-}N_e=5$, and the angular momenta $l_e^*=S_e^*=5/2$ and 
$l_{X^-}^*=S_{X^-}^*-1=3/2$.
There is one empty state in the lowest CF-$e^-$ shell giving $L_e=5/2$,
and the CF-$X^-$ has $L_{X^-}=3/2$.
Adding these two angular momenta gives $L=1$, 2, 3, and 4 as the lowest
band of $5e^-$--$X^-$ states.
The multiplicative state at $L=0$ (open circle) and the band of four 
multiplets containing an $X^-$ at $L=1$ to 4 (line) can clearly be seen 
at $d=0$ in frame (a).
Although the hidden symmetry is only approximate at $d>0$, these bands 
can be easily identified at $d=0.5\lambda$ in frame (b).

At an intermediate separation of $d=1.75\lambda$ in frame (c), neither 
description used for $d<\lambda$ or $d\gg\lambda$ is valid, and it seems
that a low energy band occurs at $L=0$, 1, 2, $3^2$, 4, 5, and 6.
Most likely, the $X^-$ unbinds but the hole is still able to bind one 
electron, forming an exciton with a significant electric dipole moment.
This dipole moment results in repulsion between the exciton the remaining 
six electrons, so that the correlations are quite different than at 
$d=0$, where the exciton decouples.

The PL spectrum can be evaluated from the eigenfunctions obtained in 
the numerical diagonalization of finite systems.
For $d\gg0$, between one and three peaks are observed in the PL 
spectrum \cite{chen2}.
Their separations are related to the Laughlin gap (for creation of
a QE--QH pair) and to the energy of interaction between the valence 
band hole and the electron system.

\subsection{
   Charged Excitons at a Finite Magnetic Field}
\begin{figure}[t]
\rule{1in}{0in}
\epsfxsize=3.75in
\epsffile{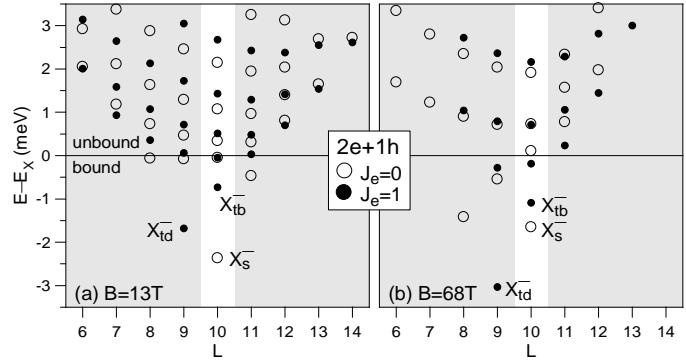}
\caption{
   The energy spectra (binding energy vs.\ angular momentum) of the 
   two-electron--one-hole system calculated on a Haldane sphere 
   with the Landau level degeneracy of $2S+1=21$.
   Five electron and hole Landau levels are included, and the parameters 
   are appropriate for the 11.5~nm GaAs quantum well in the magnetic 
   field of $B=13$~T (a) and 68~T (b).}
\label{fig16}
\end{figure}

One final point is worth mentioning.
The numerical calculations described so far were performed for 
an idealized model in which electrons and holes were confined to 
infinitely thin $2D$ layers, and only the lowest Landau level 
was considered.
For realistic systems, effects due to spin, finite width of the 
quantum well, and Landau level mixing are very important.
The energy spectra of the simple $2e$--$1h$ system calculated at 
$2S=20$ for parameters appropriate to a 11.5~nm GaAs/AlGaAs quantum 
well are shown in figure~\ref{fig16}.
Two frames correspond to the magnetic field of $B=13$~T and 68~T.
We used five electron and hole Landau levels ($n\le4$) in the 
calculation, with the realistic magnetic field dependence of the 
hole cyclotron mass and the appropriate Zeeman splittings.
The interaction matrix elements included finite (and different) 
effective widths of electron and hole quasi-2D layers.

There are a number of bound $X^-$ states in both frames, in contrast
to only one singlet bound state at $B=0$ or only one triplet bound 
state predicted for an idealized system at infinite $B$.
Three of these bound states are of particular importance.
The $X^-_s$ and $X^-_{tb}$ ($b$ for ``bright''), the lowest singlet 
and triplet states at $L=S$, are the only well bound radiative states, 
while $X^-_{td}$ ($d$ for ``dark'') has by far the lowest energy of 
all non-radiative ($L\ne S$) states.
The dark triplet state $X^-_{td}$ is the state discussed in the 
preceding sections; it is the only bound state in the lowest Landau 
level, but unbinds at low magnetic fields.
The bright singlet state $X^-_s$ is the only bound state at $B=0$, 
but unbinds at very high fields due to the hidden symmetry.
These states cross at $B\approx30$~T, as predicted in an earlier 
calculation \cite{whittaker}.
The bright triplet state $X^-_{tb}$ has been discovered very 
recently \cite{wojs4}.
It occurs only at intermediate fields and does not cross neither 
$X^-_s$ or $X^-_{td}$.
It has larger PL intensity than the $X^-_s$ state.

Although an isolated $X^-_{td}$ is non-radiative because of the
angular momentum selection rule, its collisions with other $X^-$'s 
or with electrons (which break the translational symmetry) could 
be expected to allow for $X^-_{td}$ recombination.
However, the Laughlin correlations limit high energy collisions 
at low filling density ($\nu\sim1/5$ or less) and the PL intensity 
of a dark $X^-_{td}$ remains very low also in a presence of other
particles \cite{wojs4}.
In consequence, the $X^-_{td}$ is not seen in PL, and there is no 
contradiction between experiment \cite{hayne}, which sees recombination 
of a triplet state at the energy above the singlet state up to 50~T, 
and theory \cite{whittaker}, which predicts that the lowest triplet 
state crosses the singlet at roughly 30~T.

\section{
   Summary}
\label{sec10}
We have introduced the Jain CF mean field picture and shown how the 
low lying states can be understood by simple addition of angular 
momentum.
The mean field CF picture gives the correct spectral structure not because 
of some cancellation between Chern--Simons and Coulomb interactions beyond
the mean field, but because it selects a low angular momentum subset 
of the allowed multiplets that avoids the largest pair repulsion.
The Laughlin correlations, which describe incompressible quantum fluid 
states, depend critically on the electron pseudopotential being of 
``short range'' (by which we mean that $V(L_{12})$ increases more 
quickly than $L_{12}(L_{12}+1)$).
The validity of Jain's picture also depends upon $V(L_{12})$ being of
short range.
The pseudopotential describing quasiparticles of a Laughlin condensed 
state display short range behavior only at certain values of $L_{12}$.
We have used this fact to explain why only certain states in the CF
hierarchy give rise to incompressible states of the quasiparticle fluid
(or daughter states in the hierarchy).
The pseudopotentials $V_n(L_{12})$ for higher Landau levels ($n>0$)
do not display short range behavior at all values of $L_{12}$, implying
that Laughlin-like correlations will not necessarily result at $\nu'
=2p+\nu$, where $p$ is an integer and $\nu$ is a Laughlin--Jain filling
factor.
The CF ideas have been applied successfully to multicomponent plasmas
containing different types of Fermions with the prediction of possible
incompressible fluid states for these systems.
Finally, the energy spectrum and PL of electron--hole systems can be
interpreted in terms of CF's and Laughlin correlations.

\section*{
   Acknowledgment}
The authors gratefully acknowledge the support of Grant DE-FG02-97ER45657
from the Materials Science Program -- Basic Energy Sciences of the US
Department of Energy.
They wish to thank P. Hawrylak, P. Sitko, I. Szlufarska, and K.-S. Yi
for helpful discussions on different aspects of this work.

\section*{
   References}

\end{document}